\documentclass[a4paper,11pt]{article}

\usepackage[margin=1in]{geometry}

\usepackage{physics}
\usepackage{aas_macros}

\usepackage{authblk}
\usepackage{cprotect}

\usepackage{graphicx}
\usepackage{subcaption}
\usepackage{amsmath,amssymb}
\usepackage{rotating}

\usepackage[dvipsnames,table]{xcolor}

\usepackage{hyperref}
\hypersetup{
  colorlinks   = true, 
  urlcolor     = Emerald, 
  linkcolor    = magenta, 
  citecolor    = blue 
}

\newcommand{\commander}{\texttt{Commander}}
\newcommand{\sevem}{\texttt{SEVEM}}
\newcommand{\planck}{\textit{Planck}}
\newcommand{\nside}{$N_{\rm side}$}
\newcommand{\npix}{$N_{\rm pix}$}
\newcommand{\healpix}{\texttt{HEALPix}}

\newcommand{\cellcolr}{\cellcolor{Apricot}}

\title{A reassessment of LVE method and hemispherical power asymmetry in CMB temperature data from Planck PR4}

\author[1]{Sanjeev Sanyal}
\author[1]{Sanjeet K. Patel}
\author[1]{Pavan K. Aluri\thanks{pavanaluri.phy@iitbhu.ac.in}}
\author[2,3]{Arman Shafieloo\thanks{shafieloo@kasi.re.kr}}
\affil[1]{Dept. of Physics, Indian Institute of Technology (BHU), Varanasi - 221005, India}
\affil[2]{Korea Astronomy and Space Science Institute (KASI), Yuseong-gu, 776 Daedeok daero, Daejeon 34055, Republic of Korea}
\affil[3]{University of Science and Technology (UST), Yuseong-gu 217 Gajeong-ro, Daejeon 34113, Republic of Korea}

\date{\today}

\begin{document}

\maketitle

\begin{abstract}
We undertake a reassessment of one of the large angular scale anomalies observed in cosmic microwave background (CMB) temperature signal referred to as Hemispherical Power Asymmetry (HPA). For the present analysis we use \sevem\ cleaned CMB maps from \planck's 2020 final data release (public release 4/PR4). To probe HPA, we employed the local variance estimator (LVE) method with different disc radii ranging from $0.5^\circ$ to $90^\circ$.
Our emphasis here is to revalidate the LVE method in various ways for its optimal usage and probe the hemispherical power asymmetry in the form of a dipole modulation field underlying CMB sky. By and large, our results are in agreement with earlier reported ones with more detailed presentation of explicit and not-so-explicit assumptions involved in the estimation process.
It is reaffirmed that HPA is confined to low multipoles or large angular scales of the CMB sky. A dipole like anisotropy was found in the LVE maps with anomalous power for disc radii of $2^\circ$ and upward up to $36^\circ$ at $\gtrsim2\sigma$ level. In the range $4^\circ$ to $10^\circ$ none of the 600 \sevem\ CMB simulations were found to have a dipole amplitude higher than the data when using LVE method as originally proposed.
The above reported values fall in the reliability range of LVE method after this extensive re-evaluation.
We also observe a scale dependence of the HPA dipole amplitude and model it as a power-law.
We conclude that the hemispherical power asymmetry still remains as a challenge to the standard model.
\end{abstract}

\section{Introduction}
\label{sec:intro}
Cosmological principle i.e., our universe is homogeneous and isotropic on large enough scales, is the concept on which the edifice of standard model of cosmology was built. Many tried to validate this principle using a variety of cosmological data. A powerful means to probe Cosmological principle is cosmic microwave background (CMB) radiation which was measured through ground as well as space-borne missions.
So far three full-sky CMB experiments were conducted from space that produced maps of CMB anisotropies with varying precision~\cite{Smoot1991,Bennett_2003,plk13overview}.
These missions were aimed at measuring CMB with increasing precision, particularly the temperature anisotropies, and many more are in commissioning or planning phase that will be operational in the future \cite{LiteBird_2022,PICO_2019,CMB_Bharat_Adak_2022,PIXIE_Kogut_2011,SimonsObs2019,CMBS4ScienceCase2019}. The temperature anisotropy maps (and also the polarized CMB sky, though noise dominated) from WMAP and \planck\ satellite missions were studied extensively for any hints of deviations from statistical isotropy vis-a-vis the Cosmological principle. Such studies resulted in finding various anomalies that violate the assumption of isotropy. These anomalies were found in both WMAP and \planck\ data sets indicating their robustness across two different missions with different instrument properties and systematics~\cite{wmap7yranom,wmap9yrmaps,plk2013isostat,plk2015isostat,plk2018isostat}.

From the first and three year data release of full-sky cleaned CMB maps from NASA's WMAP satellite, an excess power in one hemisphere compared to the opposite hemisphere was found that violates statistical isotropy of CMB sky. This asymmetry centered around $(l_{\rm d},b_{\rm d})=(225^\circ,-27^\circ)$ in galactic coordinates has come to be known as hemispherical power asymmetry (HPA)~\cite{Eriksen_2004,Eriksen_2007}. Since then a variety of methods in real (pixel) and harmonic (multipole) space were devised to probe its presence and robustness \cite{Hansen_2004,Hajian2005,Prunet2005,Lew2008,Bernui2008,Hoftuft_2009_lowL,HansonLewis2009,Paci2010,FlenderHotchkiss2013,Pranati2013,Akrami2014,QuartinNotari2015,Aiola_2015,Ghosh_2016,Tarun2019anom,Gimeno_Amo_2023}. Its impact on inferred cosmological parameters was studied, for example, in Refs.~\cite{Axelsson2013,Suvodip2016,YeungChu2022}.

In Ref.~\cite{Gordon2005}, a dipole modulation of otherwise isotropic CMB sky was proposed to describe the observed hemispherical power asymmetry. Thus we have,
\begin{equation}
\Delta T_{\rm obs}(\hat{n}) \equiv \Delta T_{\rm mod}(\hat{n}) = [1 +  M(\hat{n})] \Delta T_{\rm iso}(\hat{n}) = (1 +  \vec{d} \cdot \hat{n}) \Delta T_{\rm iso}(\hat{n}) 
\label{eq:cmb-dip-mod}
\end{equation}
where $\Delta T_{\rm obs}(\hat{n})$/$\Delta T_{\rm mod}(\hat{n})$ is the observed/modulated (anisotropic) CMB sky, and $\Delta T_{\rm iso}(\hat{n})$ is the otherwise isotropic CMB temperature sky.
Further, $\vec{d} = A_{\rm d} \hat{\lambda}$ where $A_{\rm d}$ and $\hat{\lambda}$ are respectively the magnitude and direction of the modulating field `$M(\hat{n})$' that is taken to be purely dipolar in nature. No modes higher than a dipole in the modulating field were seen in the data \cite{plk2013isostat,Akrami2014}.

{ The hemispherical power asymmetry anomaly is not yet resolved. In this work, we make a reassessment of the efficiency and limitation, if any, of the widely used LVE method in probing HPA (at various angular scales of the CMB sky).
There are some explicit and not-so-explicit assumptions in the procedural aspects of LVE method in testing HPA that we first appraise in this work.
Then, we propose an alternate choice for computing local variance maps (LVMs) from a CMB map which not only gives the same result but is computationally fast due to less number of patches over which local variances need to be derived.
Finally, we evaluate the scale dependence of the HPA amplitude, if any, so as to facilitate a better model building from a theoretical perspective.}

\section{Local variance estimator}
\label{sec:methods}
Following the Cosmological principle (CP), specifically isotropy, the properties of the universe inferred from different directions should be consistent with each other within error bars. For example, the number of galaxies in a given solid angle in any chosen direction or the properties of CMB anisotropies from any patch of any shape at any location should have same properties. In the standard cosmological model based on CP, CMB anisotropies are expected to be a statistically isotropic Gaussian random field.

Thus we consider a simple shape for the patch viz., a circular disc of some chosen radius defined at various locations covering the entire sky to map the variance - our chosen property of CMB anisotropy field - locally.
CMB is like a signal on a sphere and is digitized using the \healpix\footnote{\url{https://healpix.sourceforge.io/}} pixelization scheme (that is to represent data on a sphere).
In \healpix, surface of a sphere is partitioned into 12 base tiles and each tile is further divided into $N_{\rm side}^2$ pixels. So there are a total of \npix=12$\times N_{\rm side}^2$ pixels in a \healpix\ map characterized by the parameter \nside. This \nside\ parameter is chosen in powers of `$2$' as \nside$=2^m$ where $m=0,1,2,\cdots$.
Therefore, a higher \nside\ map denotes a high resolution map.
Thus, a CMB temperature anisotropy signal $\Delta T(\hat{n})$ is equivalent to pixelized map $\Delta T(p)$ where `$p$' is the pixel index whose pixel center is taken to be the direction of the incoming photon along `$\hat{n}$'.
Maps of CMB sky derived from NASA's WMAP and ESA's \planck\ satellite were usually made available at \nside=512 and 2048 respectively.

Now, we can use a \healpix\ grid of any \nside\ (typically lower than the original CMB map's resolution) to map the variances computed locally by defining circular discs of different radii `$r$'.
Let $\hat{N}$ be used to denote pixel centers of the variance map's \healpix\ grid so that the discs defined on this grid uniformly cover the entire sky.
Thus the local variance estimator (LVE) can be defined as~\cite{Akrami2014},
\begin{equation}
\sigma_r^2 (\hat{N}) =\frac{1}{N_p} \sum_{p{\in}r@{\hat{N}}} (T(p) - \bar{T}_r)^2\,,   
\label{eq:lve}
\end{equation}
where $\bar{T}_r$ is the mean value of CMB fluctuations inside the circular disc of size `$r$', and `$p$' being the pixel indices of all pixels falling within the same disc centered in the direction $\hat{N}$ of a local variance map. Further `$N_p$' are the number of pixels in the circular disc from which variance is being computed locally. For brevity, we used $T(p)$ instead of $\Delta T(p)$ to denote CMB temperature anisotropies.

Following Eq.~(\ref{eq:cmb-dip-mod}), the LVE map of a dipole modulated CMB sky, upto first order in `$A_{\rm d}$' is given by\footnote{See Eq.~7 of Ref.~\cite{rajib2023hpaai} and the text surrounding it for an explicit demonstration.},
\begin{equation}
\sigma^2_{\rm obs} (\hat{N}) \approx \sigma^2_{\rm iso} (\hat{N}) (1 + 2 A_{\rm d} \hat{\lambda} \cdot \hat{N})\,.
\label{eq:lve-dip-mod}
\end{equation}
Assuming that the amplitude of isotropy violating modulation field to be small, we considered terms only up to first order in `$A_{\rm d}$' in the above equation. So an LVE map's dipole amplitude will be \emph{twice} that of a dipole modulated CMB map to leading order. 

Thus, in order to probe the properties of the LVE map for any signatures of isotropy violation, we define a normalized variance map as,
\begin{equation}
\xi_r (\hat{N}) = \frac{\sigma_{r,\rm obs}^2(\hat{N})- \langle \sigma_{r,\rm iso}^2(\hat{N}) \rangle}{\langle \sigma^2_{r,\rm iso}{(\hat{N})} \rangle}
\label{eq:norm-lve}    
\end{equation}
where $\langle{\sigma}^2_{r,\rm iso}(\hat{N}) \rangle$ is the expected bias in an LVE map due to random fluctuations in a particular realization of CMB sky such as our universe. It is computed from simulated CMB maps based on standard cosmological model incorporating appropriate measurement artifacts such as detector noise, etc.
These simulations are also injected with anisotropic Doppler boosting and gravitational lensing effects that induce correlations between multipoles. However, for brevity, we refer to these simulations as ``\emph{isotropic}'' realizations. We also use simulation ensembles that have dipole modulation injected in them which will be discussed later.
Accordingly, the normalized LVE map of a dipole modulated CMB sky following Eq.~(\ref{eq:cmb-dip-mod}) and (\ref{eq:lve-dip-mod}) will be,
\begin{equation}
\xi (\hat{N}) \equiv 2 A_{\rm d} \hat{\lambda} \cdot \hat{N}\,.
\label{eq:norm-lve-dip-mod}
\end{equation}

However, there will always be a residual monopole along with the dipole in the normalized variance map of a given map. So,
\begin{equation}
\xi (\hat{N}) = A_m + A_{\rm LV}\hat{\lambda}\cdot \hat{N}\,,
\label{eq:norm-lve-dip-mod-param}
\end{equation}
where the expected monopole $\langle A_m \rangle=0$ and `$A_{\rm LV}$' is the LVE map's dipole amplitude. Comparing the above two equations, we get $A_{\rm LV}=2 A_{\rm d}$.

Now, there are some practical consideration in employing the LVE method.
First, irrespective of the cleaning procedure used to extract \emph{clean} CMB sky from raw satellite data, there will always be residual contamination in the cleaned CMB map thus obtained. This is particularly true in regions closer to galactic plane where foreground emission of astrophysical origin is strong. {The adopted mask} also excludes bright extended as well as point sources of extragalactic origin in the microwave sky. Hence, in order not to bias our inferences, the circular discs defined in deriving LVMs are taken in conjugation with the galactic mask omitting those pixels or regions of the sky, where the recovered CMB signal is deemed potentially contaminated. Then, to compute our estimator reliably i.e., local variances, we do not consider variances from circular discs in which 90\% or more fraction of the pixels compared to those in a full circular disc get omitted as a consequence of masking in our analysis. 

To derive LVE maps, we select discs of different sizes that allows us to probe any angular/multipole dependence of the amplitude of the underlying (dipole) modulation field. The angular scale of a CMB fluctuation, say $\theta$, is related to a multipole as $l \sim 180^\circ/\theta$. Choosing a disc radius of size `$r$' to compute local variances, we are filtering the CMB map up to a multipole of order $\ell \sim 180^\circ/r$. Correspondingly, the amplitude of modulation will be $\equiv A_{{\rm d},\ell}$. If the underlying modulation field is same at all scales, then, $A_{{\rm d},\ell}=A_{\rm d}$ (some constant to be determined from LVE maps using different disc sizes).

Now, even if the CMB sky is isotropic (in data or a simulation), there will always be a random dipole component (and other higher order modes) present in the corresponding LVE map. Thus, following Eq.~(\ref{eq:cmb-dip-mod}), we have :
\begin{equation}
\langle d_i \rangle = 0 \quad {\rm and} \quad \langle d_i^2 \rangle \neq 0\,, \quad \forall {\quad} i=x,y,z\,,
\label{eq:dipole-expt-val}
\end{equation}
for the components of the dipole in the modulating field.
Since the amplitude is given by,
\begin{equation}
A_{\rm d} = \sqrt{d_x^2 + d_y^2 + d_z^2}\,,
\label{eq:dipole-A}
\end{equation}
assuming only a dipole-like component in the modulating field, such a quantity is always non-zero.
By a simple transformation one can show that the dipole power, say, $C_1$ is related to the dipole amplitude, $A_{\rm d}$, as $C_1 = 4\pi A_{\rm d}^2/9$. Therefore, to get the \emph{correct} dipole power underlying the CMB sky, one should subtract expected random dipole power, $\langle C_{1,\rm iso} \rangle$, which is like a noise term i.e.,
$C_{1,\rm corr} = C_{1,\rm obs}-\langle C_{1,\rm iso} \rangle$ or equivalently,
\begin{equation}
A^2_{\rm d,corr} = A^2_{\rm d,obs} - \langle A^2_{\rm d,iso} \rangle\,.
\label{eq:A_corr}
\end{equation}
However, the usual practice now~\cite{Akrami2014} is to make such a correction at the amplitude level i.e., $A_{\rm d,corr} = A_{\rm d,obs} - \langle A_{\rm d,iso} \rangle$ which is incorrect. 
If one is employing multipole space estimators, the \emph{unbiased} estimate of the dipole amplitude will be given by
\begin{equation}
A_{\rm d,corr}= 1.5\sqrt{(C_{1,\rm obs}-\langle C_{1,\rm iso} \rangle)/\pi}\,.
\end{equation}

In order to estimate the dipole in an LVE map we use \healpix\ function \verb+remove_dipole+ which returns the dipole amplitude, `$A_{\rm d}$', as well as its direction $(\theta_{\rm d},\phi_{\rm d})$ (or equivalently $(l_{\rm d},b_{\rm d})$) in galactic coordinates which is the coordinate system of the input CMB map used in the present study.

Here, we briefly describe how \verb+remove_dipole+ function works.
Let $m(p)$ be a map in \healpix\ format. Then the monopole and dipole of $m(p)$ are separated by the \texttt{remove\_dipole} subroutine following the decomposition:
\begin{equation}
m(p) = f_0 + \vec{f}\cdot\hat{p} +m'(p) = f_0 + f_1 x +f_2 y + f_3 z + m'(p) = \sum_{j=0}^{d^2-1} f_j s_j(p) + m'(p)\,,
\end{equation}
where $f_0$,  $\vec{f}$ and $m'(p)$ are the monopole, dipole and rest of map that contains multipoles $l\geq2$. Here $\hat{p}$ are the sky coordinates of the pixel center indexed as `$p$' and $\{s_j(p)\}=(1,x,y,z)$ for $j=0,\cdots,3$. Further, $d=1,2$ if one wants to fit only monopole or both monopole and dipole, respectively.

The fit parameters $f_0$ and $\vec{f}$ are obtained by minimizing the weighted square of the residuals,
\begin{equation}
    \Sigma^2 (\{f_i\}) = \sum_{p\in D} w(p) \, \left[ m(p)-\sum_{j=0}^{d^2-1} f_j s_j(p)\right]^2\,,
\end{equation}
with respect to the fit parameters $\{f_i\}$ which result in a linear system of equations,
\begin{equation}
\sum_{j=0}^{d^2-1} A_{ij} f_j = b_i\,,
\end{equation}
to solve, where,
\begin{equation}
A_{ij}=\sum_{p\in D}s_i(p)w(p)s_j(p) \quad {\rm and} \quad b_i=\sum_{p\in D}s_i(p)w(p)m(p)\,.
\end{equation}
The summation here is over the pixels `$p$' from the allowed region `$D$' of the sky. If there is no masking, then `$D$' will be the entire sky. Further, $w(p)$ are the weights with which one wants to fit a particular pixel `$p$' in estimating `monopole' or `monopole + dipole'. If we don't pass any weights, then \verb+remove_dipole+ function of \healpix\ will take $w(p)=1$ for all pixels being used in the fit. More details can be found in \healpix\ documentation\footnote{\url{https://healpix.sourceforge.io/documentation.php} where {\tt FORTRAN90} subroutines are described}.

\section{Data and Complementary simulations}
\label{sec:data-sim}

\begin{figure}
\centering
\includegraphics[width=0.48\textwidth]{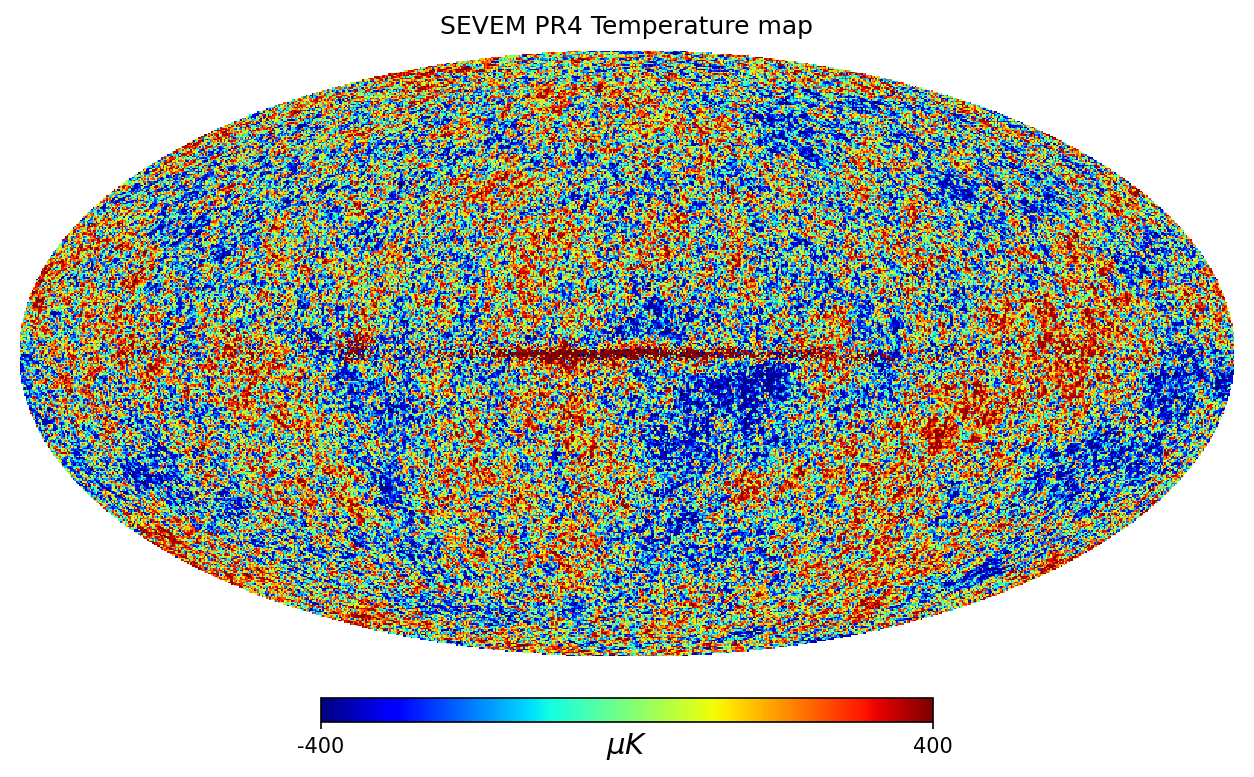}
~
\includegraphics[width=0.48\textwidth]{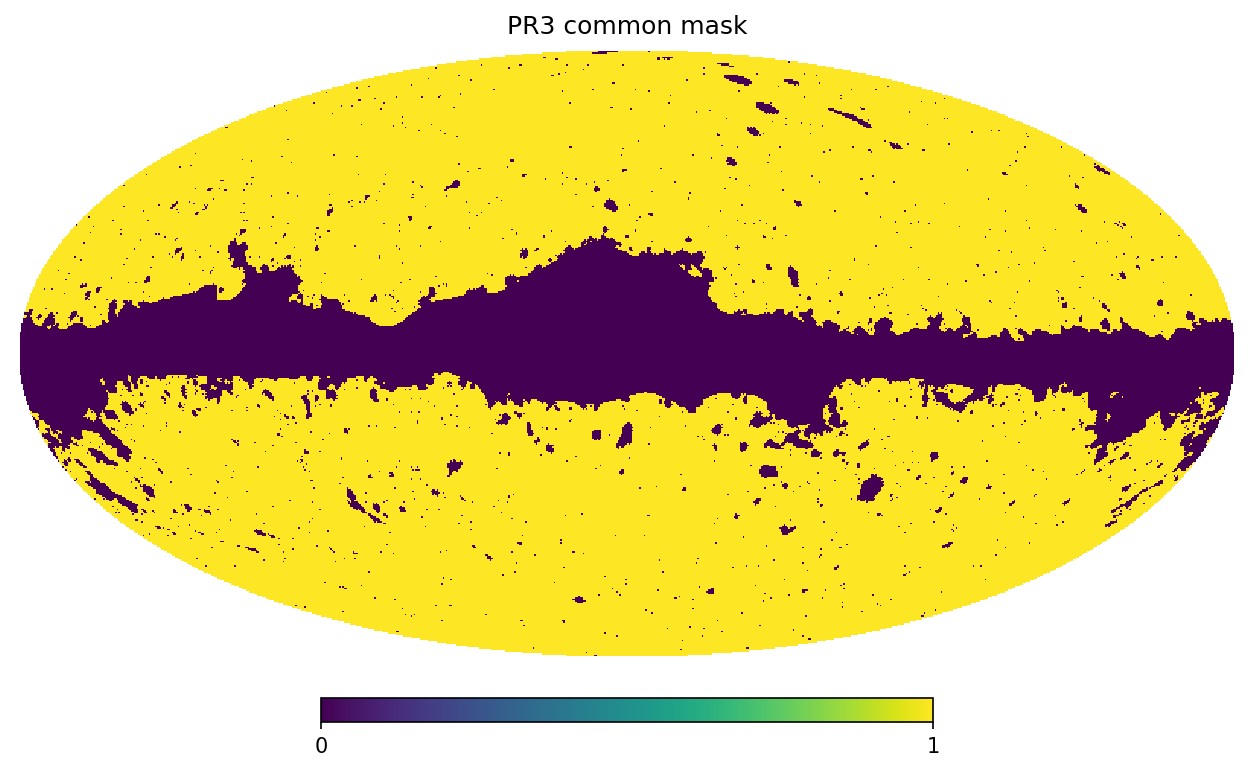}
\caption{\emph{Left} : The \sevem\ cleaned CMB map from \planck's PR4 data release. 
\emph{Right} : \planck\ PR3 \emph{common} mask recommended for use with PR4 cleaned CMB maps.}
\label{fig:sevem-mask-maskedsevem}
\end{figure}

\subsection{Observed CMB sky and galactic mask used}
From \planck's public release 4 (PR4) data~\cite{npipe2020}, we use a cleaned CMB map that was derived using the \sevem\ cleaning procedure~\cite{leach2008,sevem2012}. The \sevem\ cleaning procedure is one of the variants of the most commonly used internal linear combination (ILC) methods in real (pixel) space. It involves forming foreground ``templates'' by taking differences of observed raw satellite maps (in thermodynamic units) from specific frequency channels that serve as a proxy for one or more of the foregrounds present in the microwave sky. As a result of such subtraction, the difference maps don't have any CMB signal in them. These difference maps are then subtracted with appropriate weights from rest of the observed maps (at other frequencies that are not used in forming foreground ``templates'') to get a \emph{cleaned} CMB map. For more details regarding the \sevem\ method, the reader may consult the original papers cited above. \commander\ processed PR4 CMB temperature map is also available along with 100 simulations at \nside=4096. However, we do not consider \commander\ map in our present analysis due to fewer number of complementary simulations.

The \sevem\ cleaned CMB map from the entirety of \planck's observations in different (a total of nine) frequency bands is produced at \nside=2048 with a detector beam resolution given by a Gaussian window function of FWHM$=5'$ (arcmin). No matter how good a cleaning algorithm may be, in regions of strong galactic emission, the recovered CMB signal is not reliable and hence omitted from any kind of cosmological analysis. In order to omit such regions, \planck\ collaboration provided a confidence mask along with the CMB map derived using \sevem\ procedure, as is the case with any other cleaning procedure. However a combined mask referred to as \emph{common} mask was also provided by \planck\ collaboration for use with any of the cleaned CMB maps that it produced. These masks are provided as a binary map with $\{0,1\}$ where pixels/regions that should be omitted are set to `$0$' and the allowed pixels/regions are set to `$1$'. The common mask has, expectedly, lower non-zero sky fraction, often denoted by  the symbol $f_{\rm sky}$, compared to individual cleaning method (component separation) specific CMB confidence masks.
We use the PR3 common mask~\cite{plk18CompSep}, as recommended, that has an available sky fraction of $f_{\rm sky} \approx 77.94\%$.

In Fig.~[\ref{fig:sevem-mask-maskedsevem}], we show the \sevem\ cleaned CMB map from \planck's PR4 and the recommended \planck's PR3 \emph{common} mask for temperature analysis of PR4 data.

\subsection{Simulations}
Among the suite of data products released by \planck\ as part of PR4, simulations corresponding to \sevem\ CMB maps are also provided that are referred to as FFP (Full Focal Plane) simulations, specifically the FFP12 set of CMB realizations along with noise. CMB and noise maps are simulated at the nine frequency channels in which \planck\ made observations with appropriate beam and pixel window effects. A total of 600 sets of nine frequency specific CMB and noise maps were provided as part of \planck\ PR4. This is less than the usual number of 1000 simulations that were provided with each of the earlier \planck\ public releases. Each of these simulation sets have nine channel maps that are combined using the same weights that were applied on data to do foreground reduction. Thus, a total of 600 simulated CMB and noise maps are provided separately mimicking observations. These have the same beam and \healpix\ resolution as the data viz., a beam resolution given by a Gaussian of FWHM$=5'$ and \nside=2048.
These are made available on NERSC facility\footnote{\url{http://crd.lbl.gov/cmb-data}} and \planck\ legacy archive\footnote{\url{https://pla.esac.esa.int/}} (PLA).

LVE method is validated and used in several studies including \planck\ team's tests of isotropy and statistical consistency of CMB sky with standard model expectations based on Cosmological principle \cite{plk2015isostat,plk2018isostat}. However we validate the method again, presenting some of the not so explicitly demonstrated aspects of LVE methodology.

We used three sets of CMB realizations in our study. The first set is the 600 \sevem\ FFP12 simulations corresponding to \emph{isotropic} random realizations of CMB sky based on standard cosmological model. As mentioned earlier, though they have anisotropic signals such as Doppler boosting and lensing effects included, we refer to these as \emph{isotropic} maps for brevity and also in the sense that they are ``unmodulated'' maps. Their spherical harmonic coefficients are given by $a_{lm}^{\rm iso} = b^{5'}_l p^{2048}_l a_{lm}^{\rm CMB}$, where $b^{5'}_l$ denotes the $5'$ (arcmin) Gaussian beam transfer function and $p^{2048}_l$ is the pixel window function corresponding to \healpix\ \nside=2048 due to the finite size of the pixels. From these isotropic realizations, we generate the following two sets of dipole modulated simulations for re-validating the LVE method.

The second set of simulations are dipole modulated maps at all scales with a fixed amplitude along a particular direction as previously found in the data. To generate these \emph{pure-dm} simulations, first we get the spherical harmonic coefficients, $a_{lm}^{\rm iso}$, of FFP12 CMB mock maps using \texttt{map2alm} routine of \healpix. These $a_{lm}^{\rm iso}$ are then deconvolved with corresponding beam smoothing ($b^{5'}_l$) and pixel window function ($p^{2048}_l$) to apply a dipole modulation (per Eq.~\ref{eq:cmb-dip-mod}) with an amplitude $A_{\rm d}=0.072$, directed along $(l_{\rm d},b_{\rm d}) = (221^\circ, -20^\circ)$ in galactic coordinates \cite{plk2018isostat} on the deconvolved CMB map. Finally, we convolve this modulated CMB map with the same beam and pixel window functions ($b_l^{5'}$ and $p_l^{2048}$). Thus we obtain dipole modulated CMB maps with the same amplitude at all scales using the \texttt{alm2map} routine of \healpix.

A third set of simulations were also generated to simulate the observed case of dipole modulation in data, where the modulation is present only on large angular scales of the CMB sky up to $l\sim 64$ \cite{Hoftuft_2009_lowL,HansonLewis2009}.
In order to simulate such \emph{low-l-dm} maps, we combine the spherical harmonic coefficients of an FFP12 \emph{isotropic} \sevem\ CMB map and a \emph{pure-dm} map with the same index using a ``cosine'' filter defined in multipole space for smoothly transitioning between dipole modulated $a_{lm}$'s at low multipoles to isotropic $a_{l m}$'s at high multipoles. The filter is defined as,  
\begin{equation}
f_{l} = \left\{
\begin{array}{ll}
1 & l < l_{1} \\ 
\frac{1}{2} \left[ 1 + \cos \left( \pi \frac{l - l_1}{l_2 - l_1} \right) \right] & l_1 \leq l \leq l_2 \\
0  & l > l_2
\end{array}\right.
\end{equation}
choosing $l_1 = 60$ and $l_2 = 70$ for our purpose. Thus a \emph{low-l-dm} CMB map with dipole modulation confined to large angular scales is generated as,
\begin{equation}
a_{l m}^{\rm low-l-dm} = f_l \, a_{l m}^{\rm pure-dm} + (1-f_l) a_{l m}^{\rm iso}
\end{equation}
where $a_{l m}^{\rm low-l-dm}$, $a_{l m}^{\rm pure-dm}$, $a_{l m}^{\rm iso}$ are the spherical harmonic coefficients of low multipole dipole modulated, all scale dipole modulated, and isotropic CMB realizations respectively. These $a_{l m}$'s are then converted back to pixel domain (using \texttt{alm2map} routine of \healpix).

Finally, all these three sets of CMB realizations are added with FFP12 noise maps having the same index to get the final noisy (an)isotropic CMB realizations. Note that the nine FFP12 \emph{noise} maps corresponding to nine frequency channels of a particular simulation (index) were also combined by the \planck\ collaboration in a way similar to the  \sevem\ FFP12 CMB realizations using the same data derived weights for foreground cleaning.

We also note that few works have used \planck\ PR4 data for probing HPA among other features~\cite{Gimeno_Amo_2023,Aghanim2024sz}. However, their emphasis were different and the LVE method was used as it was originally proposed~\cite{Akrami2014}. Here, we are reassessing both the LVE procedure, as well as the data (viz., \planck\ PR4) in light of this revalidation of the LVE method.

\section{Analysis and Results}
\label{sec:anls-rslts}

\subsection{Validation}
\label{sec:vald}

Here we present a (re)validation of the LVE method for probing the dipole modulation signal supposedly underlying the observed CMB sky.

In the first application of local variance estimator method \cite{Akrami2014}, a \healpix\ grid of \nside=16 was used whose pixel centers uniformly cover the entire sky to map the locally computed variances. One would want to use temperature anisotropy information (pixel values of digitized CMB map) from disjoint regions so that they are uncorrelated with each other. However, depending on the size of the disc radius, the locally computed variances will have overlapping pixels from adjoining regions when using a fixed \nside\ grid (=16, in this case), making them correlated. 
It is also possible that the circular discs used to compute variances locally are smaller than the pixel size of the chosen \healpix\ grid of the LVE map itself and doesn't make use of all the pixels in original CMB map (that is usually at a higher resolution, which is at \nside=2048 in our case compared to \nside=16 of the LVE map).

One resolution to this situation is to choose the region covered by each pixel of a lower \nside\ \healpix\ grid as mask to map local variances, by upgrading the LVE map's grid with one pixel at a time to that of the input CMB map. This will keep the regions disjoint and the locally computed variances uncorrelated in estimating the underlying dipole modulation signal. However as shown in Table~\ref{tab:hpx-pixsize}, this can be done only for a few select disc radii that matches with the pixel size of LVE map's \nside. These pixel sizes are estimated assuming the \healpix\ pixels to be a square of side $\sqrt{4\pi/N_{\rm pix}}\times 180^\circ/\pi$ in degrees, where $4\pi$ is the area of the entire spherical surface (of a unit sphere) that is divided into \npix\ pixels using \healpix.

If one wishes to undertake a detailed study of HPA as a function of angular scale (or equivalently as a function of multipole i.e., to study $A_\ell$), then LVE maps are derived for different choices of disc radii covering the entire CMB sky.
Other than those listed in Table~\ref{tab:hpx-pixsize}, since local variances computed from any other choice/combination of \nside\ and disc radius `$r$' will have overlapping regions, correlations between them will be unavoidable. However, to minimize the correlations in an LVE map, it would be useful to define a \healpix\ grid in deriving LVE maps that have different \nside\ so as to match the chosen disc radius with an LVM's pixel size as much as possible. In Table~\ref{tab:vary-nside-radii}, we show the different \nside\ values chosen to match the disc radius used to compute an LVE map in the present work, instead of using a fixed \nside=16, say. It also allows to probe anisotropy at smaller scales by using a suitably higher \nside\ grid for an LVE map for probing signals such as the Doppler boosting of CMB~\cite{Adhikari2015}, whose signature is evident readily at small angular scales, as a result of coupling between adjacent multipoles.

\begingroup

\renewcommand{\arraystretch}{1.2} 
\begin{table}
\centering
\begin{tabular}{| c | c | c | c |}
\hline
\nside\ & \npix=$12\times N_{\rm side}^2$ & Pixel size ($PS^\circ$) & Disc size ($r^\circ$) \\
\hline
64 & 49152 & $0.92^\circ$ & $0.5^\circ$ \\
32 & 12288 & $1.83^\circ$ & $1^\circ$ \\
16 & 3072 & $3.66^\circ$ & $2^\circ$ \\
8 & 768 & $7.33^\circ$ &  $4^\circ$ \\
4 & 192 & $14.66^\circ$ & $8^\circ$ \\
2 & 48 & $29.32^\circ$ &  $16^\circ$ \\
1 & 12 & $58.63^\circ$ & $32^\circ$ \\
\hline
\end{tabular}

\caption{Pixel size ($PS=\sqrt{4\pi/N_{\rm pix}}\times 180^\circ/\pi$ in degrees) corresponding to various \healpix\ resolutions characterized by the \nside\ parameter and the approximate disc radii ($r \sim PS/2$) that could be used with the chosen \healpix\ grid for LVE maps to match its pixel sizes.}
\label{tab:hpx-pixsize}
\end{table}

\endgroup

\begin{figure}
\centering
\includegraphics[width=0.98\textwidth]{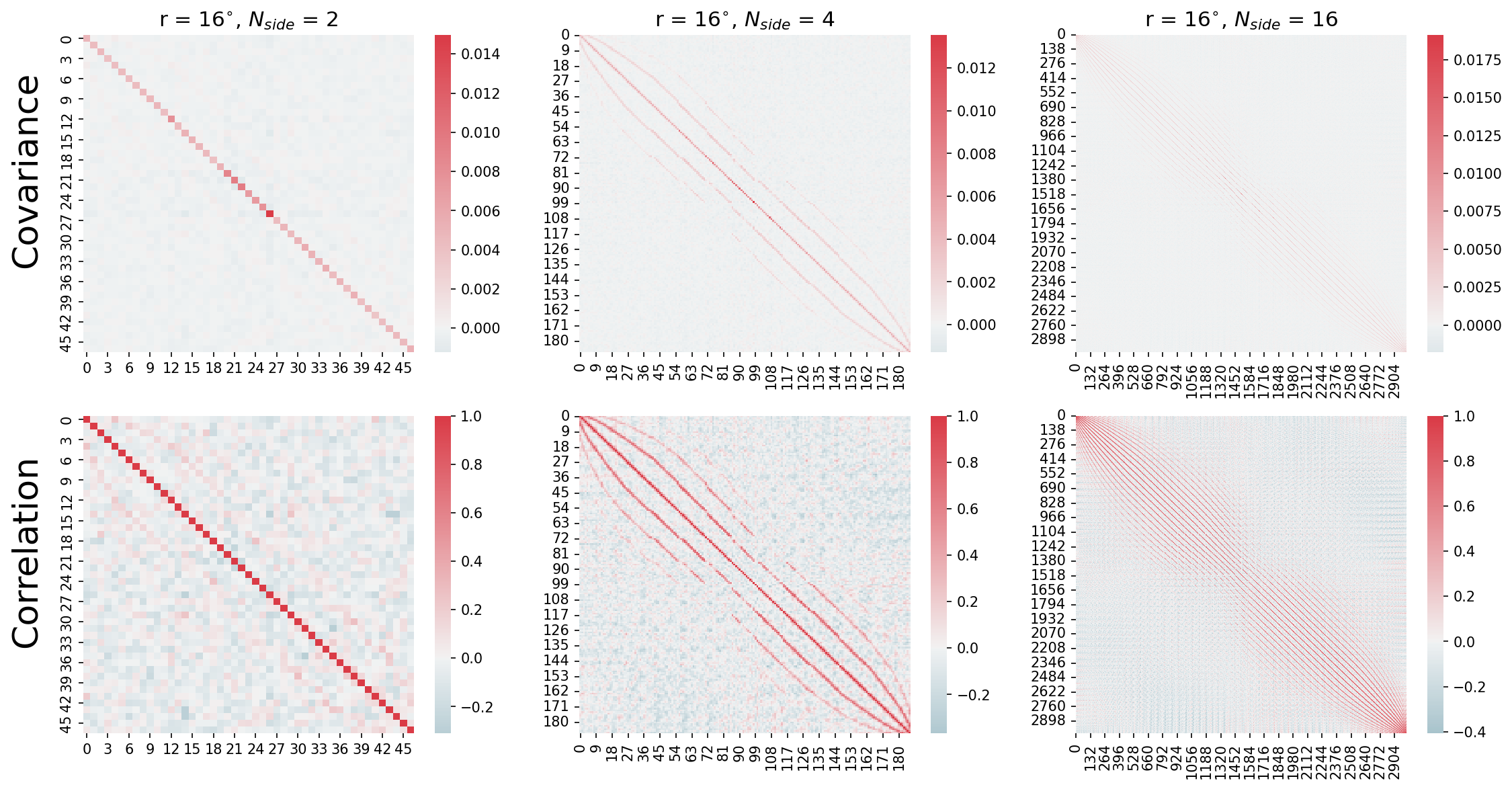}
\caption{\emph{Top}: Covariance matrices $C_{pp'}$ as defined in Eq.~(\ref{eq:cov-mat}) corresponding to normalized LVE maps generated at \nside=2, 4, and 16 but for a particular disc radius $r=16^\circ$.
\emph{Bottom}: Correlation matrices, $\tilde{C}_{pp'}$, (normalized covariance matrices) as
defined in Eq.~(\ref{eq:corr-mat}) corresponding to the same disc radius but from LVE maps with  different \nside\ as shown in the \emph{top} panels.}
\label{fig:cov-corr-matrix}
\end{figure}

Let $p$ and $p'$ be any two pixel indices of a normalized LVE map, $\xi_r(\hat{N})$, estimated for some choice of disc radius `$r$'. Then the covariance and the corresponding correlation matrix are defined in the usual way as,
\begin{eqnarray}
C_{pp'} &=& \langle \xi_{r,{\rm iso}}(p) \xi_{r,{\rm iso}}(p')  \rangle - \langle \xi_{r,{\rm iso}}(p)  \rangle  \langle \xi_{r,{\rm iso}}(p') \rangle\,, \label{eq:cov-mat} \quad {\rm and} \\
\tilde{C}_{pp'}  &=& \frac{C_{pp'}}{\sqrt{C_{pp}C_{p'p'}}}\,.
\label{eq:corr-mat}
\end{eqnarray}
The correlation matrix $\tilde{C}_{pp'}$ is defined such that the covariance matrix elements are normalized with respect to the diagonal elements, which are dominant and positive. The correlation matrix elements consequently lie in the range `$-1$' and `$+1$'. Here $\xi_{r,{\rm iso}}$ stands for normalized LVE maps from the FFP12 isotropic (\sevem) CMB realizations added with noise (per Eq.~\ref{eq:norm-lve}). In Fig.~[\ref{fig:cov-corr-matrix}], we show the covariance among pixels of normalized LVE map ($C_{pp'}$) estimated for a disc radius of $r=16^\circ$ but at different \healpix\ resolutions of \nside\ = 2, 4, 16 in the \emph{top} panel. In the same figure, the corresponding correlation matrices ($\tilde{C}_{pp'}$) are shown in the \emph{bottom} panel to highlight the relative strength of the covariance matrix elements.

From Fig.~[\ref{fig:cov-corr-matrix}], it is clear that an LVE map's \nside\ should be chosen such that its pixel size matches with the disc radius used to compute the variances of a CMB map locally. In doing so we can keep the LVE map's covariance matrix predominantly diagonal, if not diagonal, as is obvious from the first two panels (top or bottom) of Fig.~[\ref{fig:cov-corr-matrix}].
Otherwise, the covariance matrix will be a band diagonal matrix due to overlapping information in LVM pixels, as can be seen from the third panel (top or bottom) of Fig.~[\ref{fig:cov-corr-matrix}], that have to be accounted for in estimating the underlying modulation (or any anisotropy) field. For larger disc radii, we fix \nside=2 to estimate LVE maps which is sufficient and also computationally fast.
Thus we use different \nside\ grids to compute LVE maps for different choices of disc radii as listed in Table~\ref{tab:vary-nside-radii}, to keep the correlations across LVM pixels minimal. This exercise also points out that the variances of an LVE map's pixels are different as can be seen from the diagonals of covariance matrices shown in \emph{top} panels of Fig.~[\ref{fig:cov-corr-matrix}]. It is indeed known that the variance of a Gaussian random variable follows a $\chi^2$ distribution. So they will be different.
Also, the variances computed will differ slightly if the sample sizes (number of pixels used) differ. Thus we use the variances of LVM's pixels as weights (inverse variance weighting) to better estimate the dipole modulation field (using the \verb+remove_dipole+ functionality of \healpix).

\begingroup

\renewcommand{\arraystretch}{1.2} 
\begin{table}
\centering
\begin{tabular}{| r | c | c | c |}
\hline
Disc Radius ($r^\circ$)  & \nside\  & Pixel Size, $PS^\circ$ & $\sqrt{2}\times PS^\circ$ \\
\hline
0.5$^\circ$ & 64 & 0.92$^\circ$ & 1.30$^\circ$ \\
1$^\circ$ & 32 & 1.83$^\circ$ & 2.59$^\circ$ \\
2$^\circ$ & 16 & 3.66$^\circ$ & 5.18$^\circ$ \\
4$^\circ$,6$^\circ$ & 8  & 7.33$^\circ$ & 10.36$^\circ$ \\
8$^\circ$,10$^\circ$,12$^\circ$,14$^\circ$ & 4 & 14.66$^\circ$ & 20.73$^\circ$ \\
16$^\circ$,18$^\circ$,20$^\circ$,24$^\circ$,28$^\circ$,32$^\circ$,36$^\circ$, & 2 & 29.32$^\circ$ & 41.46$^\circ$ \\
40$^\circ$,50$^\circ$,60$^\circ$,70$^\circ$,80$^\circ$,90$^\circ$     &    &       & \\
\hline
\end{tabular}
\caption{\nside\ used to derive LVE maps for various choices of disc sizes `$r$'.
		Treating each \healpix\	pixel to be a square, its size (side of the square) is
		given by $PS=\sqrt{4\pi/N_{\rm pix}}$ (radians) where
		$N_{\rm pix}=12\times N_{\rm side}^2$.
		Size of the pixels as measured along the diagonal of such a square is
		$\sqrt{2}\times PS$. These numbers help us in choosing appropriate \healpix\ grid
		to generate LVE maps for a particular disc radius with minimal to no overlaps
		in the LVMs.}
\label{tab:vary-nside-radii}
\end{table} 

\endgroup

These covariance matrices were obtained using the 600 \sevem\ ``isotropic'' CMB realizations including noise provided as  part of \planck\ PR4. For each disc radius, we estimated the covariance matrix from the normalized LVE maps derived from these realizations. They were then used in the (re)evaluation of the LVE method using ``pure-dm'' and ``low-l-dm'' CMB realizations.
We remark that, from simulations, the distribution of the dipole modulation field's amplitudes or directions derived using the \verb+remove_dipole+ function of \healpix\ as-it-is differs nominally from those derived using the same function with diagonal elements of the covariance matrices, $C_{pp'}$, as weights to fit the dipole. We show some representative plots in Fig.~[\ref{fig:apdx:norm-inv-wght-dip-fit-varyns}] and [\ref{fig:apdx:norm-inv-wght-dip-fit-ns16}] in Appendix~\ref{apdx:norm-inv-wght-dip-fit}. The empirical distribution of recovered dipole amplitudes and directions from LVE maps appear to be essentially same in both the cases where a fixed \healpix\ grid of \nside=16 is used for all disc radii or a different \nside\ grid was chosen for different choices of disc radius. We think that this may be due to the high SNR of CMB temperature anisotropy measurements.
In the rest of the paper, having demonstrated that the variance of different pixels in an LVE map are different, we use diagonal elements from the covariance matrices as weights in computing the dipole component from an LVM, which is procedurally correct.

\begin{figure}
\centering
\includegraphics[width=0.98\columnwidth]{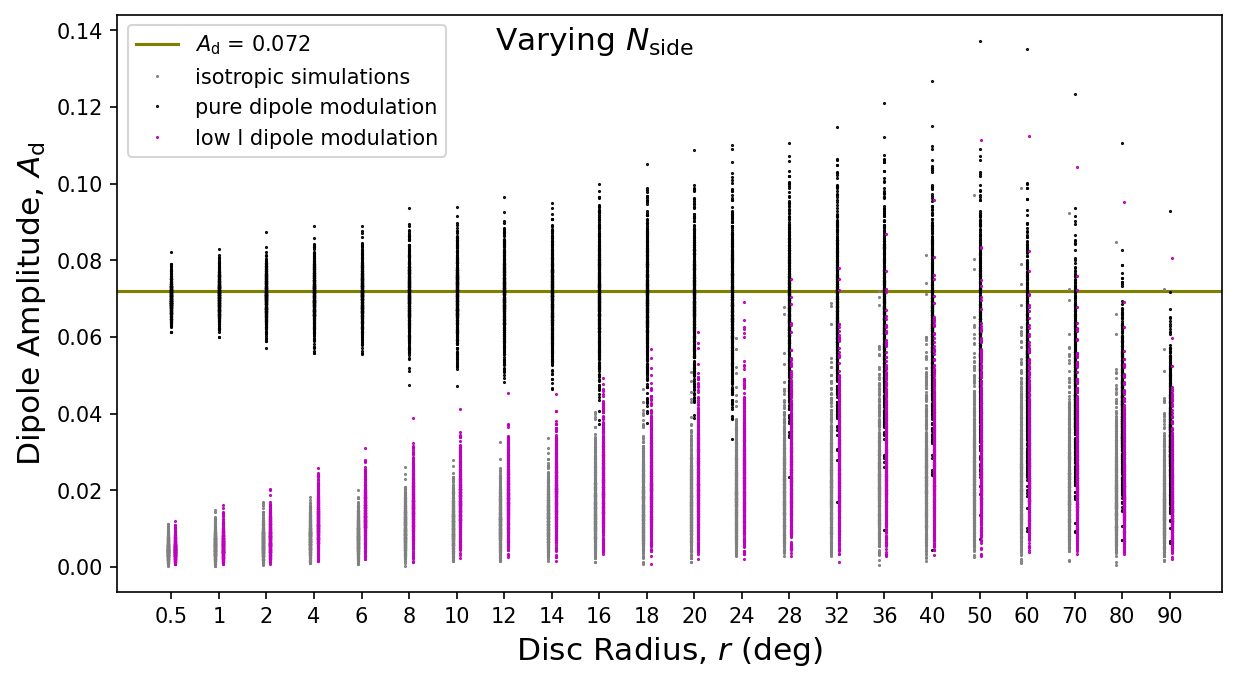}
~
\includegraphics[width=0.98\columnwidth]{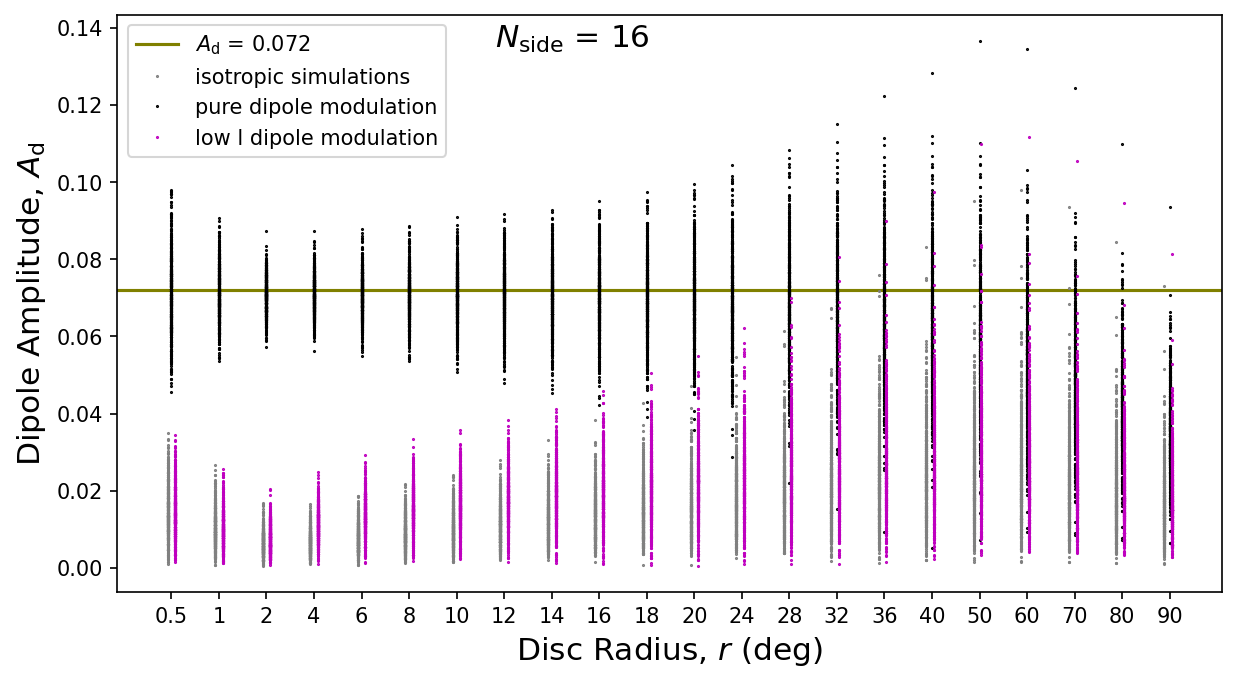}
\caption{\emph{Top} and \emph{Bottom} panels depict the distribution of dipole amplitudes,
		$A_{\rm d}=A_{\rm LV}/2$, evaluated from normalized local variance maps obtained
		with varying \nside\ and fixed \nside=16 schemes, respectively.
		Recovered dipole amplitudes from isotropic simulations are shown in \emph{gray},
		from low-$l$ dipole modulated maps in \emph{magenta} and pure dipole modulated
		maps in \emph{black}. 		
		}
\label{fig:dip-ampl-sim-varyns-ns16}
\end{figure}

\begin{figure}
\centering
\includegraphics[width=0.98\textwidth]{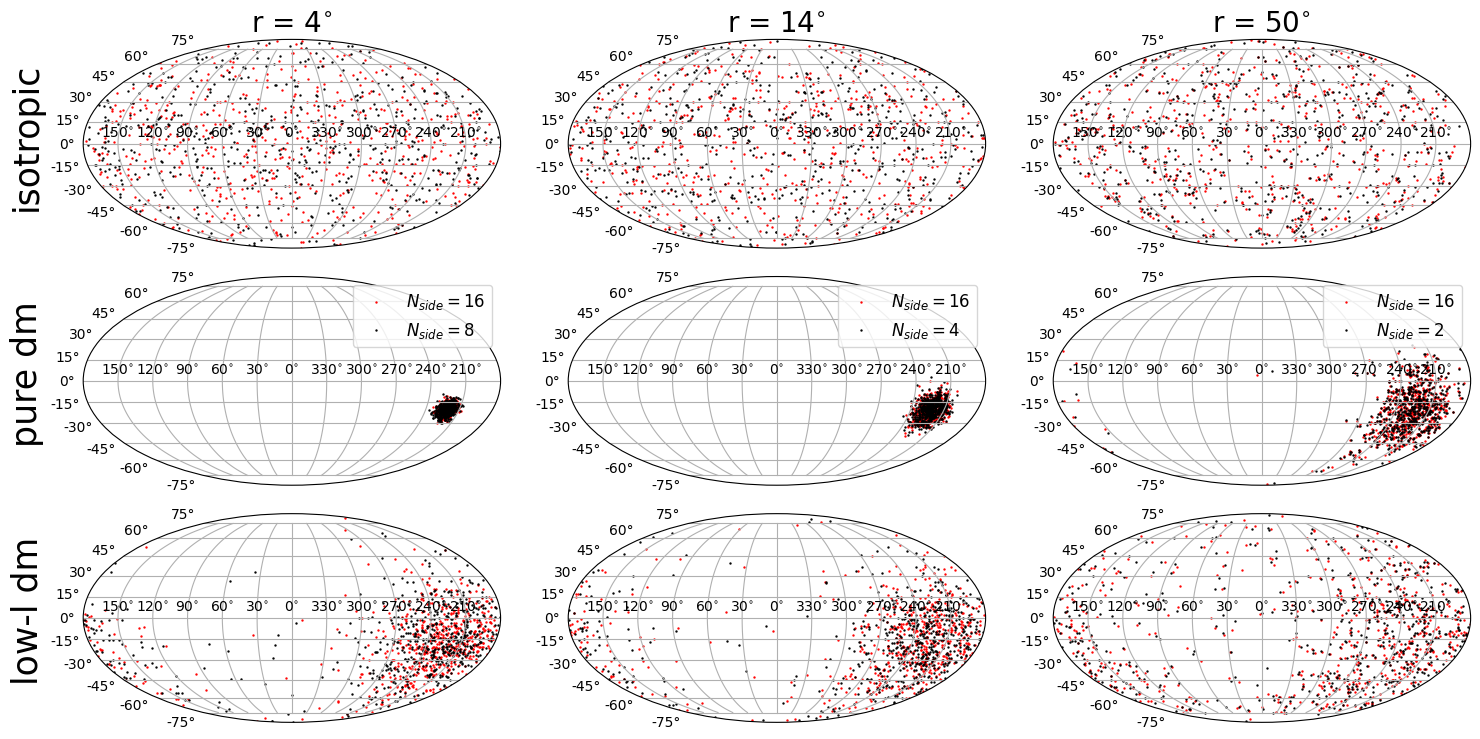}
\cprotect\caption{Recovered dipole modulation directions in LVE maps for some select disc
         radii as indicated. The \emph{top}, \emph{middle} and \emph{bottom} rows
         comparatively show the LVE maps' dipole directions
         from isotropic, pure-dm and low-$l$-dm simulation sets using
         fixed \nside=16 (\emph{red} dots) and varying \nside\ (\emph{black} dots)
         schemes. These dipoles in LVE maps are fitted using inverse variance
         weighting with the \verb+remove_dipole+ functionality of \healpix.}
\label{fig:dip-dir-3sim-varyns-ns16}
\end{figure}

From the three simulation ensembles, we present the recovered dipole amplitudes in Fig.~[\ref{fig:dip-ampl-sim-varyns-ns16}] for the different disc radii listed in Table~\ref{tab:vary-nside-radii}. { Dipole information (amplitude and direction) is derived from normalized variance maps as defined in Eq.~(\ref{eq:norm-lve})}. We used the inverse variance weighting to compute the dipole amplitude.
It is clear from the figure that the distribution/scatter in the recovered amplitudes are similar in both cases of fixed and varying $N_{\rm side}$, except for low radii where they are visibly different.
We expect that this would lead to a different $p$-value profile as a function of disc radius between the two scenarios of fixed \nside=16 and varying \nside\ LVE maps. 
The corresponding dipole directions in LVE maps, for some select disc radii, from the three different simulation ensembles created are shown in Fig.~[\ref{fig:dip-dir-3sim-varyns-ns16}]. The mean of the recovered dipole is in agreement with the injected dipole modulation field from pure-dm maps.

Now, we note some observations from these results about the LVE method.
\begin{itemize}
\item
 From Fig.~[\ref{fig:dip-ampl-sim-varyns-ns16}] (as also presented originally in Ref.~\cite{Akrami2014}), the recovered dipole amplitude of hemispherical power asymmetry is lower than the injected dipole amplitude for larger disc radii. We again show Fig.~[\ref{fig:dip-ampl-sim-varyns-ns16}] in an alternate way in Fig.~[\ref{fig:pure-dm_err_bar}]. In this figure, mean of the recovered dipole amplitudes for different `$r$' from the fixed and varying \nside\ cases are shown together.
For $r\gtrsim16^\circ$, the mean of the recovered dipole amplitudes (visually) start to deviate (towards lower values) from the injected dipole amplitude of $A_{\rm d}\approx0.07$ though the error bars also increase. This indicates a breakdown of the LVE method.

\item
Specifically, for large disc radii, the assumption of dipole being the largest angular scale feature on the sky as a background in Eq.~(\ref{eq:lve}) and (\ref{eq:lve-dip-mod}) fails, and the expected variance $\sigma_{\rm iso}^2$ cannot be taken out (as is the case for smaller disc radii compared to the dipole of the modulation field). Thus, since we know the injected value of the dipole amplitude ($A_{\rm d}\sim0.07$) in simulations, we quote the recovered dipole amplitude values from mock realizations as $A_{\rm d}=A_{\rm LV}/2$ (per Eq.~\ref{eq:norm-lve-dip-mod} and \ref{eq:norm-lve-dip-mod-param}). However, when presenting our results from data analysis, we only quote $A_{\rm LV}$ due to the breakdown of Eq.~(\ref{eq:lve-dip-mod}).

\item
{
In ``low-l-dm'' simulations, we applied the scale invariant modulation up to $l = 60$, after which the modes are isotropic. So, low-$l$-dm simulations contain effects of dipole modulation up to angular scales of $\theta \gtrsim 180^\circ/60 = 3^\circ$. 
Therefore, for disc radii $r\lesssim 3^\circ$, we find that the distribution of dipole modulation amplitudes recovered is similar to what we recover from isotropic realizations, as expected.
For $r>3^\circ$, even if there is dipole modulation signal in the simulations, angular scales contributing to the variance computed are mostly isotropic modes. Thus, although the distribution of LVE maps' dipole amplitudes shift slightly upward (in Fig.~[\ref{fig:dip-ampl-sim-varyns-ns16}]), it is however nominal. With larger disc radius compared to $r=3^\circ$, the same trend continues as isotropic CMB modes act as noise in the estimation of dipole modulation signal and also perhaps due to increasing cosmic variance at large angular scales. Finally, beyond some disc radius ($r\sim40^\circ$ if $A_{\rm d}\sim 0.07$), the assumptions of LVE methodology are not valid and thus lead to similar results as found using isotropic or pure-dm realizations.

Similar trend is seen in the recovered dipole \emph{directions} from low-$l$-dm simulation ensemble where the scatter is more compared to the same recovered from pure-dm simulations.}

\item
Various horizontal lines in Fig.~[\ref{fig:pure-dm_err_bar}] are useful in inferring a cutoff disc size for employing LVE method depending on one's tolerance level in estimating the dipole amplitudes. If one deems a $5\%$ decrement in the mean of the recovered dipole amplitudes (i.e., $\sim 0.95A_{\rm d}$) as tolerable with respect to the injected dipole amplitude, while being consistent { within the standard deviation obtained from the simulations that is used represent error bars in the plot}, a maximum disc radius of $\sim 28^\circ$ can be employed where the \emph{black} dash-dotted line crosses the mean dipole amplitude curve. If one can tolerate a $10\%$ and $20\%$ decrement i.e., $0.9A_{\rm d}$ and $0.8A_{\rm d}$ indicated by a \emph{brown} dotted line and a \emph{green} dashed line respectively in the mean recovered dipole amplitude recovered, but nevertheless contains the injected dipole amplitude `$A_{\rm d}$' within error bars, then we can go up to a disc size of $\sim40^\circ$ and $\sim 55^\circ$ respectively.

\item
The error bars on the recovered dipole amplitude increases with increasing disc radius where the assumption in Eq.~(\ref{eq:lve-dip-mod}) breaks down. So we should use the LVE method to probe isotropy violation underlying CMB sky such as HPA up to only some maximum disc radius to get a reliable estimate. Also note that these uncertainties folds in the large cosmic variance associated with low multipoles. The red/blue circle point types joined by solid lines along with error bars shown in Fig.~[\ref{fig:pure-dm_err_bar}] are the mean and the standard deviation (specifically, 16th and 84th quantile values - although they are not very different) respectively as derived from the 600 \sevem\ processed PR4 CMB realizations.

\item
Initially, we analyzed LVE maps from $r=1^\circ$ to $90^\circ$ disc radii. However to understand the patterns that were just noted more clearly at small disc radii, we extended our analysis down to $r=0.5^\circ$. We find that the uncertainty in recovered dipole amplitudes increase further at low disc radii when using fixed \nside=16 \healpix\ grid to probe HPA. This is contrary to what is seen when using varying \nside\ grid to derive LVE maps. It is true even in the case of \emph{pure-dm} simulations as can be seen from Fig.~[\ref{fig:dip-ampl-sim-varyns-ns16}] and [\ref{fig:pure-dm_err_bar}]. So, indeed employing varying \nside\ grid is the right strategy. This could be a result of completely disconnected regions in the sky being mapped via LVMs when the disc radii are smaller than the pixel size of an \nside=16 grid ($\sim 3.66^\circ$ from Table~\ref{tab:hpx-pixsize}).

\begin{figure}[t]
   \centering
   \includegraphics[width=0.98\columnwidth]{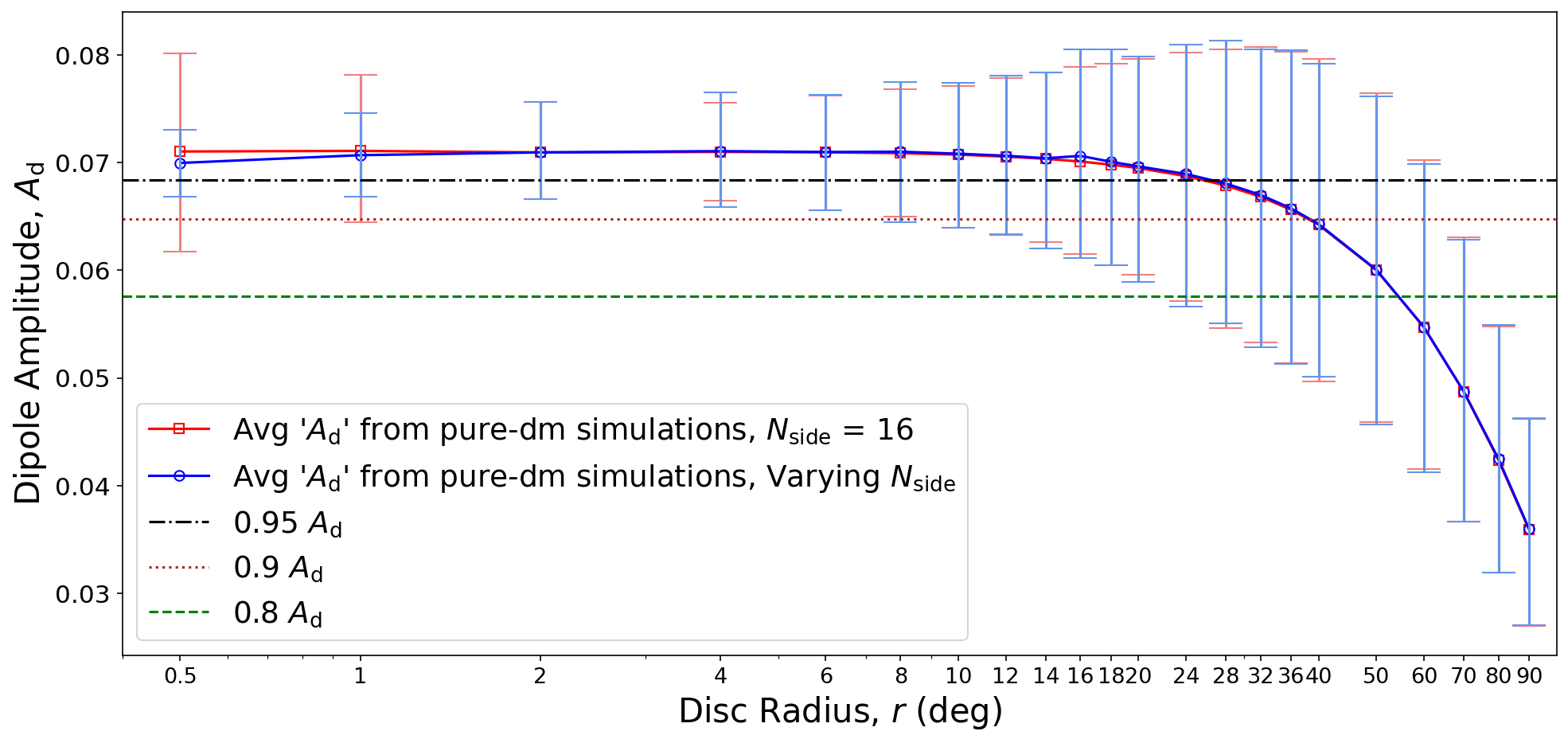}  
   \caption{Mean dipole amplitude recovered from LVE maps of pure-dm simulation ensemble with corresponding standard deviation as error bars are presented for different choices of disc size `$r$'. LVM computed with a fixed \nside=16 and varying \nside\ are shown in \emph{red} square and \emph{blue} circle point types. The \emph{black}, \emph{brown} and \emph{green} horizontal lines correspond to different tolerance levels, specifically, $5\%$, $10\%$ and $20\%$ decrement in recovered dipole amplitude compared to injected amplitude of $A_{\rm d}=0.072$.
   Wherever these lines  cross the mean dipole amplitude curve, they indicate the disc radius beyond which the LVE method may not be appropriate for that tolerance level.}
   \label{fig:pure-dm_err_bar}
\end{figure}

\begin{figure}[t]
   \centering
  \includegraphics[width=0.98\columnwidth]{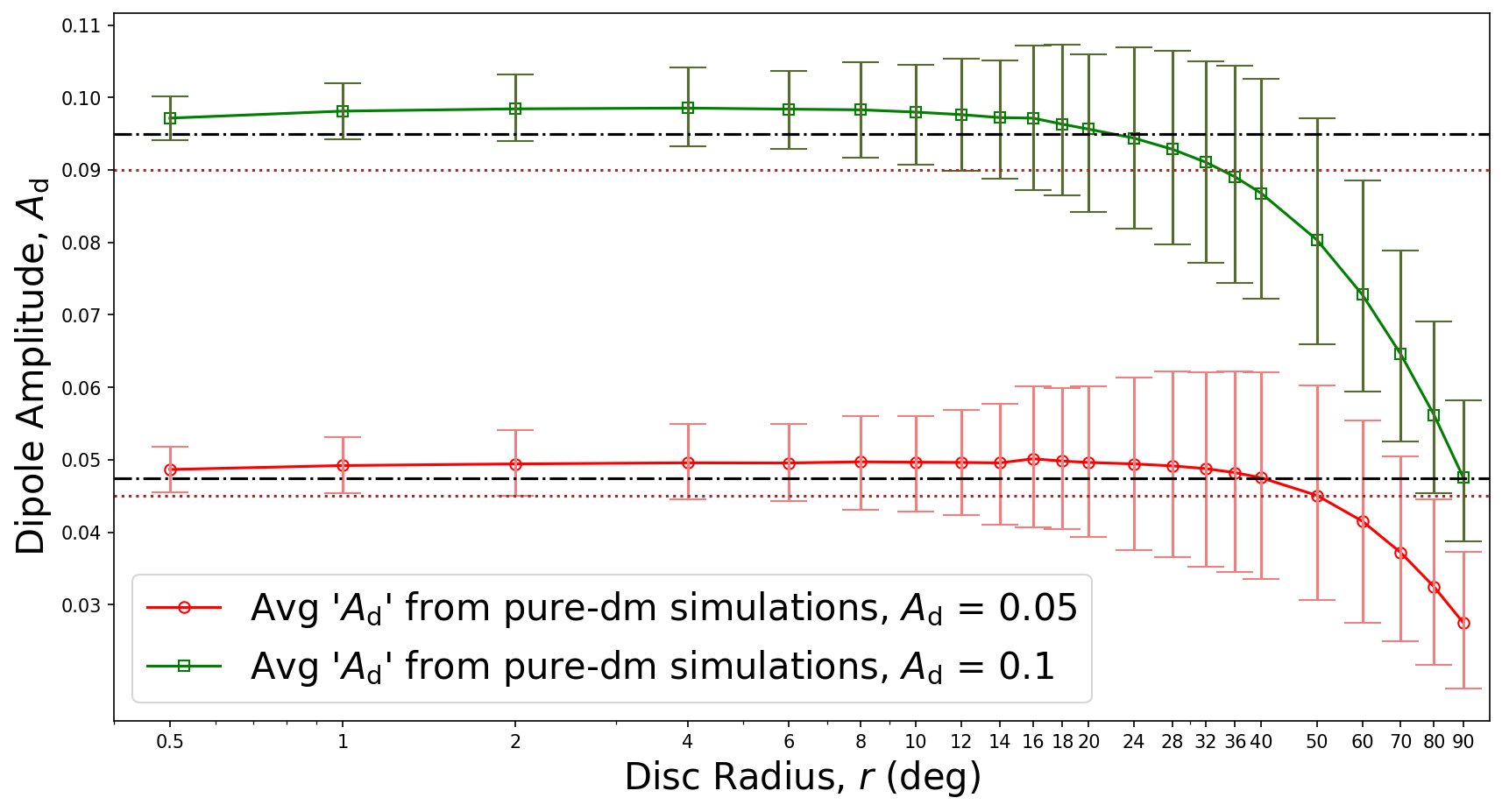}  
   \caption{Evaluating the maximum disc radius that can be chosen to use LVE method for
   different levels of anisotropy i.e., dipole modulation amplitudes of $A_{\rm d}=0.1$
   and $0.05$ underlying a CMB sky. The \emph{black} dash-dotted and \emph{brown} dotted horizontal lines indicate tolerance thresholds corresponding to $5\%$ and $10\%$ lower amplitude relative to the injected values of $A_{\rm d}=0.1$ and $0.05$, respectively.
   We clearly see a dependence in the LVE method's reliability range of disc radii on the strength of anisotropy.
   For a 10\% admissible level of deviation of mean recovered dipole amplitude with
   respect to the injected value, but however consistent within the standard deviation obtained from simulations (=error bars in the plot), LVE
   method can be employed choosing a maximum disc radius of $r\approx34^\circ$
   and $50^\circ$ for $A_{\rm d}=0.1$ and $0.05$, respectively.
   Only pure-dm CMB realizations with noise were used here.
   Also, for this illustration, the LVE maps were derived using the varying \nside\
   approach only.}
   \label{fig:pure-dm_err_bar_diff_A}
\end{figure}

\item
Here one can ask a followup question as to whether the maximum disc radius of reliability of LVE method is dependent on the level of anisotropy? To address this, we simulated dipole modulated (pure-dm) simulations with two other choices of dipole amplitudes viz., $A_{\rm d}=0.05$ and $A_{\rm d}=0.1$. For this exercise, LVE maps with varying \nside\ were computed from these ``pure-dm'' realizations. The results are presented in Fig.~[\ref{fig:pure-dm_err_bar_diff_A}]. As is obvious from the figure, for lower levels of isotropy violation, the LVE method is reliable up to a higher disc radius compared to stronger levels of isotropy violation where the method breaks down (Eq.~\ref{eq:lve-dip-mod} and \ref{eq:norm-lve-dip-mod}) at a relatively lower disc radius. Specifically, for $A_{\rm d}=0.05$ the LVE method holds up to a choice of $r\approx 50^\circ$,  whereas for an anisotropy level of $A_{\rm d}=0.1$ it is reliable up to only $r\approx 34^\circ$ to probe HPA with a tolerance level for  $10\%$ deviation in the mean recovered dipole amplitude from simulations, but however consistent with the injected dipole amplitude {within the standard deviation computed from simulations that are used to depict error bars in the plot}.

\item 
From Fig.~[\ref{fig:pure-dm_err_bar}], we also note that the effects of beam (which is uncorrected for in real space estimators) starts to show up in the recovered dipole amplitudes from LVE maps derived at very low disc radii. The mean recovered dipole amplitude at $r=0.5^\circ$ is slightly lower than the same at higher `$r$'. 

\end{itemize}

With this extensive re-evaluation of LVE method using simulations from \planck\ PR4, we proceed with presentation of results from analyzing the data. We note that using varying \nside, not only do we get to compute local variances from nearly non-overlapping regions of the CMB sky under study, but also at reduced computational cost.

\subsection{Data analysis and Results}

\begin{figure}
\centering
\includegraphics[width=0.98\textwidth]{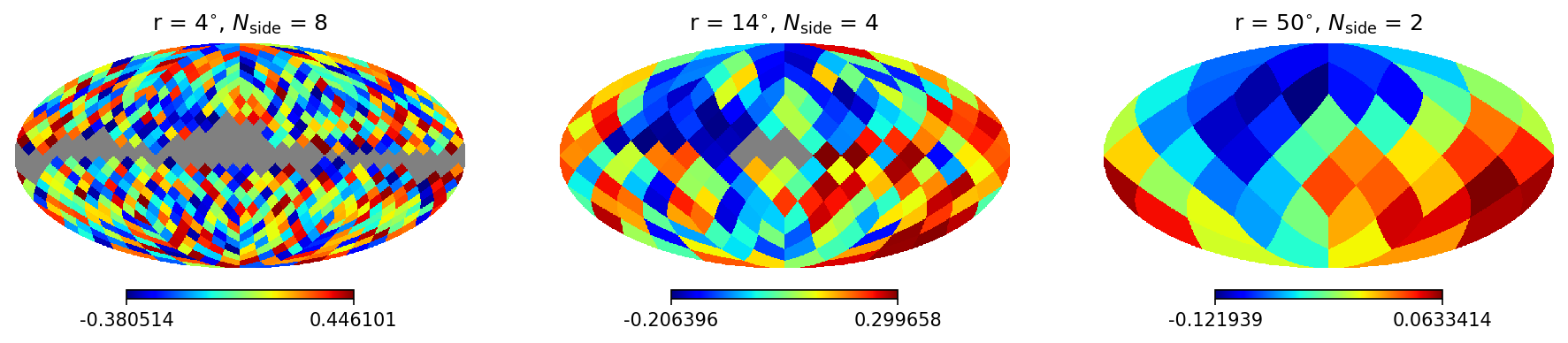}
~
\includegraphics[width=0.98\textwidth]{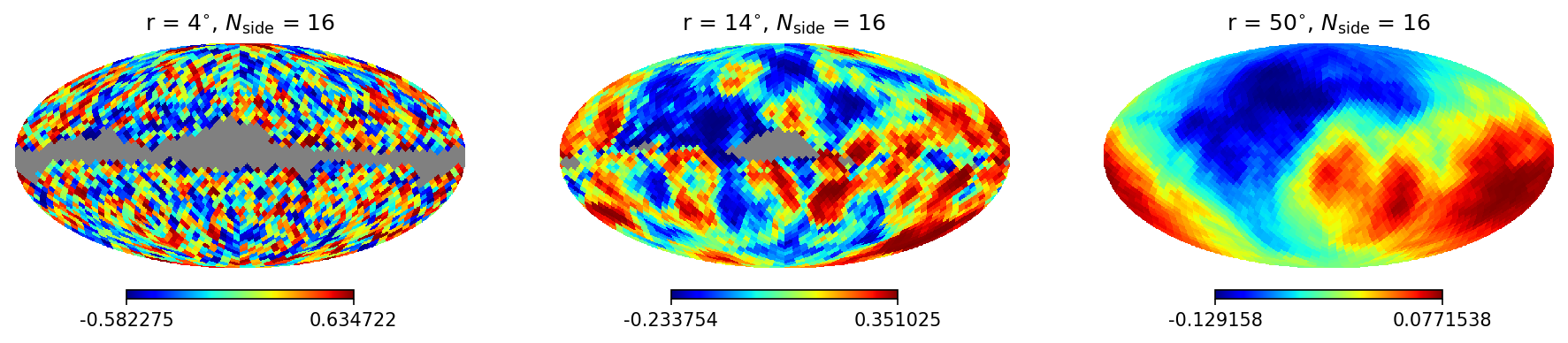}
\caption{Normalized LVE maps (per Eq.~\ref{eq:norm-lve})
         corresponding to \sevem\ CMB solution from \planck\ PR4  for some select disc
         radii are shown here. In the \emph{top} row, LVE maps obtained using varying
         \nside\ grid are shown, and in the \emph{bottom} row the same are shown
         for fixed \nside=16 grid in deriving LVE maps.
         From \emph{left} to \emph{right} the LVE maps shown correspond to
         disc radii $r=4^\circ$, $14^\circ$, and $50^\circ$.}
\label{fig:lve-varyns-ns16}
\end{figure}

As already mentioned in the previous section, we will be using the diagonal elements of the covariance matrices, derived using FFP12 isotropic simulation ensemble from \planck\ PR4,  as weights for inverse variance weighting to fit the dipole in LVE maps. We use the \verb+remove_dipole+ function of \healpix\ package to do the dipole fitting.
Further, we will be presenting results from both the cases where a fixed \nside=16 and varying \nside\ \healpix\ grid are used to derive LVE maps from data/simulations, and make comparative statements. Note that, following Eq.~(\ref{eq:norm-lve-dip-mod}), $A_{\rm LV}=2A_{\rm d}$ where the LVE methodology is reliable. The same relation holds for $A_{\text{LV,corr}}$ regarding bias correction as it is for $A_{\rm d,corr}$ following Eq.~(\ref{eq:A_corr}).

In Fig.~[\ref{fig:lve-varyns-ns16}], LVE maps from \planck\ PR4 \sevem\ cleaned CMB sky for some select disc radii are shown. Here, we chose LVE maps corresponding to disc radii $r=4^\circ$, $14^\circ$, and $50^\circ$ for illustration. Both cases of varying \nside\ to match different choices of disc radii (\emph{top} row) and fixed \nside=16 (\emph{bottom} row) for LVMs are shown in the figure. Note that these are normalized variance maps as defined in Eq.~(\ref{eq:norm-lve}).

\begin{figure}
\centering
\includegraphics[width=0.98\columnwidth]{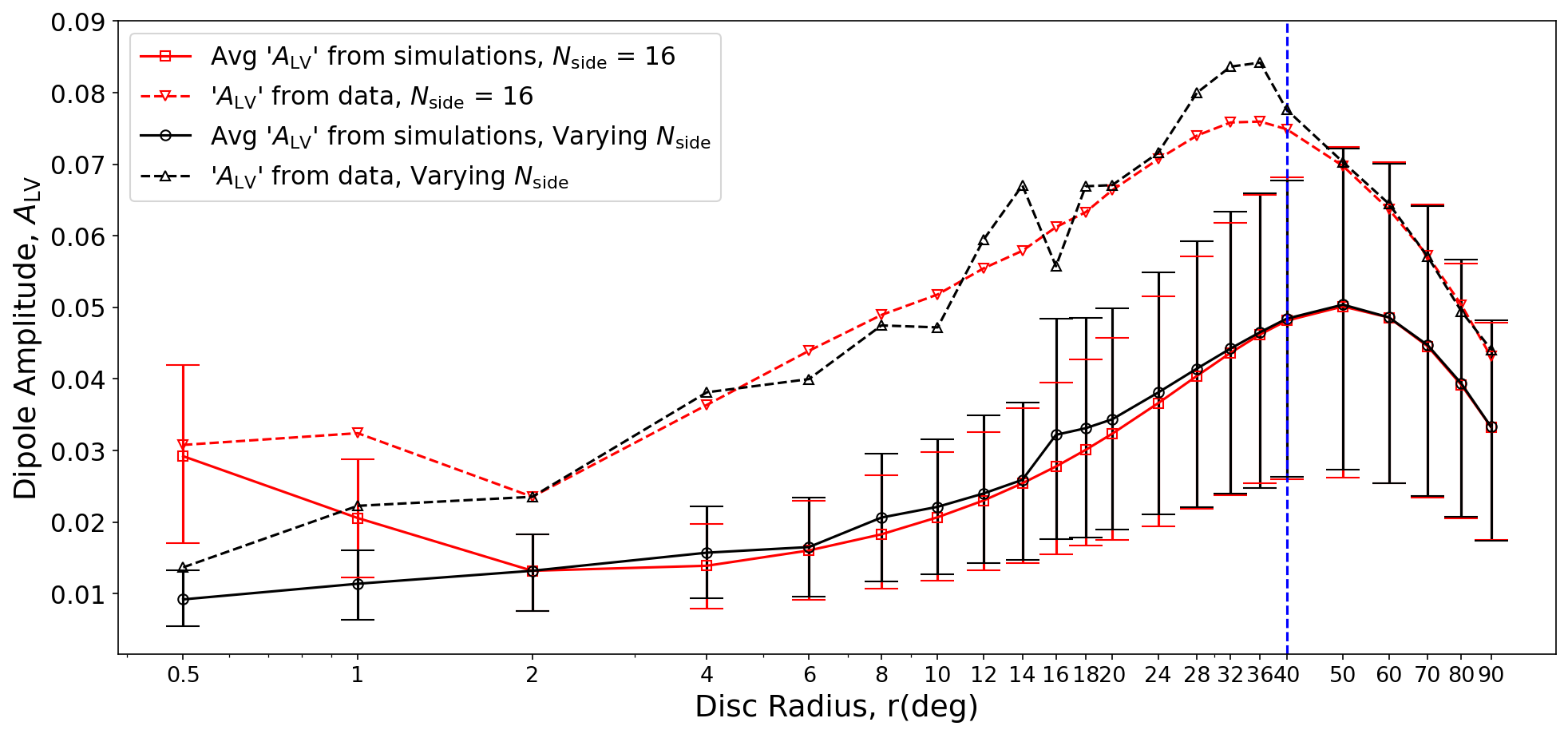}
~
\includegraphics[width=0.98\columnwidth]{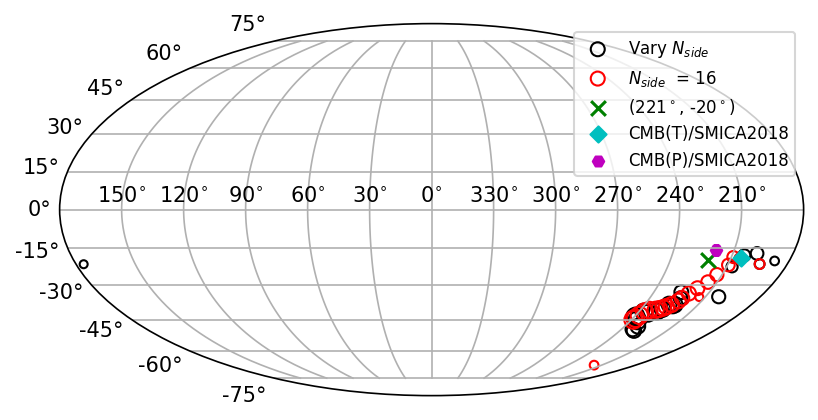}
\caption{\emph{Top:} Dipole amplitudes from data, and mean with $1\sigma$ error bars from isotropic simulations when using a fixed \nside=16 (\emph{red} - inverted triangle/square) and varying \nside\ ({\emph{black} - triangle/circle}) grids to generate normalized
variance maps, respectively, are shown. The dipole amplitudes from isotropic simulations (\emph{black} circle and \emph{red} square point-types joined by solid lines), and data  values (\emph{black} triangle and \emph{red} inverted triangle point-types joined by dashed lines) are presented as they are obtained from LVE maps (i.e., $A_{\rm LV}$).
\emph{Bottom:} Dipole directions extracted from data maps in same coloured open circle point types are shown here.  The varying circle sizes correspond to the different disc radii employed in the local variance analysis viz., larger/smaller circle sizes correspond to larger/smaller disc radii respectively.
\sevem\ CMB map from \planck\ PR4 is used as input data. Additionally, the \emph{green} cross, \emph{cyan} diamond and \emph{purple}
hexagon points denote the dipole direction used in generating the simulations in the present work, LVE map's dipole from SMICA 2018 CMB temperature map, and E-mode polarization map respectively.}
\label{fig:data-dip-ampl-error-ds}
\end{figure}

Our central findings are presented in Fig.~[\ref{fig:data-dip-ampl-error-ds}]. In the \emph{top} panel of the figure, we show the dipole amplitudes from LVE maps, $A_{\rm LV}$, estimated for different disc radii ranging from $r=0.5^\circ$ to $90^\circ$.
Both cases of using a fixed \nside=16 grid (in \emph{red} colour with inverted triangle point type) and a varying \nside\ grid (in \emph{black} colour with triangle type) to obtain LVE maps are shown in that figure.
The mean dipole amplitude, $A_{\rm LV}$, with $1\sigma$ error bars (16th and 84th quantile) from LVMs derived from ``isotropic'' \sevem\ FFP12 CMB mocks with noise are shown in \emph{red} colour with square point type and \emph{black} colour with circle point type for fixed \nside=16 and varying \nside\ choices, respectively.
One can readily see that the amplitude of the dipole component of LVE maps from data always lies outside the  $1\sigma$ error bars from simulations over a large range of disc radii chosen to compute the LVE maps.    

The estimated dipole amplitude from the two cases of fixed \nside=16 and varying \nside\ are visibly different, for $r\lesssim40^\circ$ in data. From simulations this difference vanishes for $r\gtrsim32^\circ$ (as evident from \emph{red} and \emph{black} solid lines in Fig.~[\ref{fig:data-dip-ampl-error-ds}] though consistent with each other within the standard deviation of the simulations, represented by error bars in the plot in the entire range of disc radii used).
As noted earlier due to large uncertainty in the recovered LVE maps' dipole amplitudes at $r\lesssim2^\circ$ in all three simulation ensembles when employing the fixed \nside=16 grid, the predicted dipole amplitudes underlying LVMs are higher than the actual levels. On the face value, such higher amplitude values would imply a higher level of isotropy violation. But from data, we find that the LVM dipole amplitudes are smaller/consistent with expectations at these lower disc radii. Hence, there may be no HPA present at those angular scales suggesting a scale dependent dipole modulation of the CMB sky.

Beyond $r\gtrsim50^\circ$ (in data), there is no distinction between the mean dipole amplitudes of LVE maps whether one uses fixed \nside=16 or varying \nside\  grid to map local variances. For $r\gtrsim50^\circ$, the disc radii are much larger than the pixel size of the LVE map's chosen \nside\ grid that it makes no difference in the recovered dipole amplitude between the two cases (see Table~\ref{tab:vary-nside-radii}). The vertical \emph{blue} dashed line in the top panel of Fig.~[\ref{fig:data-dip-ampl-error-ds}] indicates the reliability range of the LVE method in probing HPA for the observed level of anisotropy.

In the \emph{bottom} panel of Fig.~[\ref{fig:data-dip-ampl-error-ds}], the dipole directions recovered from LVE maps for the same choice of disc radii are shown. The open circles in \emph{red} colour denote dipole directions estimated from LVE maps using fixed \nside=16 grid and that in \emph{black} colour are same for varying \nside\ case. They remarkably span the same region in the sky with nearly overlapping directions. The dipole directions estimated from the data LVE maps with varying \nside\ have a similar trend as those from fixed \nside = 16 data LVE maps except for few radii.
Although, they collectively show a similar trend/span, when we compare the individual directions recovered for a particular disc radius from the two cases, we find that they are different. One such noticeable difference is the dipole directions for $r = 1^\circ$ estimated from the two cases that are $\sim50^\circ$ apart. For a smaller disc radius of $r=0.5^\circ$, they are $\sim 36^\circ$ away from each other.
As noted earlier, from the top panel of Fig.~[\ref{fig:data-dip-ampl-error-ds}], there is a clear dissimilarity in the estimated  dipole amplitudes for smaller disc radii in LVE maps. Also indicated in bottom panel of Fig.~[\ref{fig:data-dip-ampl-error-ds}] is the injected dipole direction~\cite{plk2018isostat} as a \emph{green} cross used in the production of modulated realizations.
Further shown in the same figure are the HPA directions reported by \planck\ collaboration as found in PR3 data~\cite{plk2018isostat}.

\begin{figure}
   \centering
   \includegraphics[width=0.98\columnwidth]{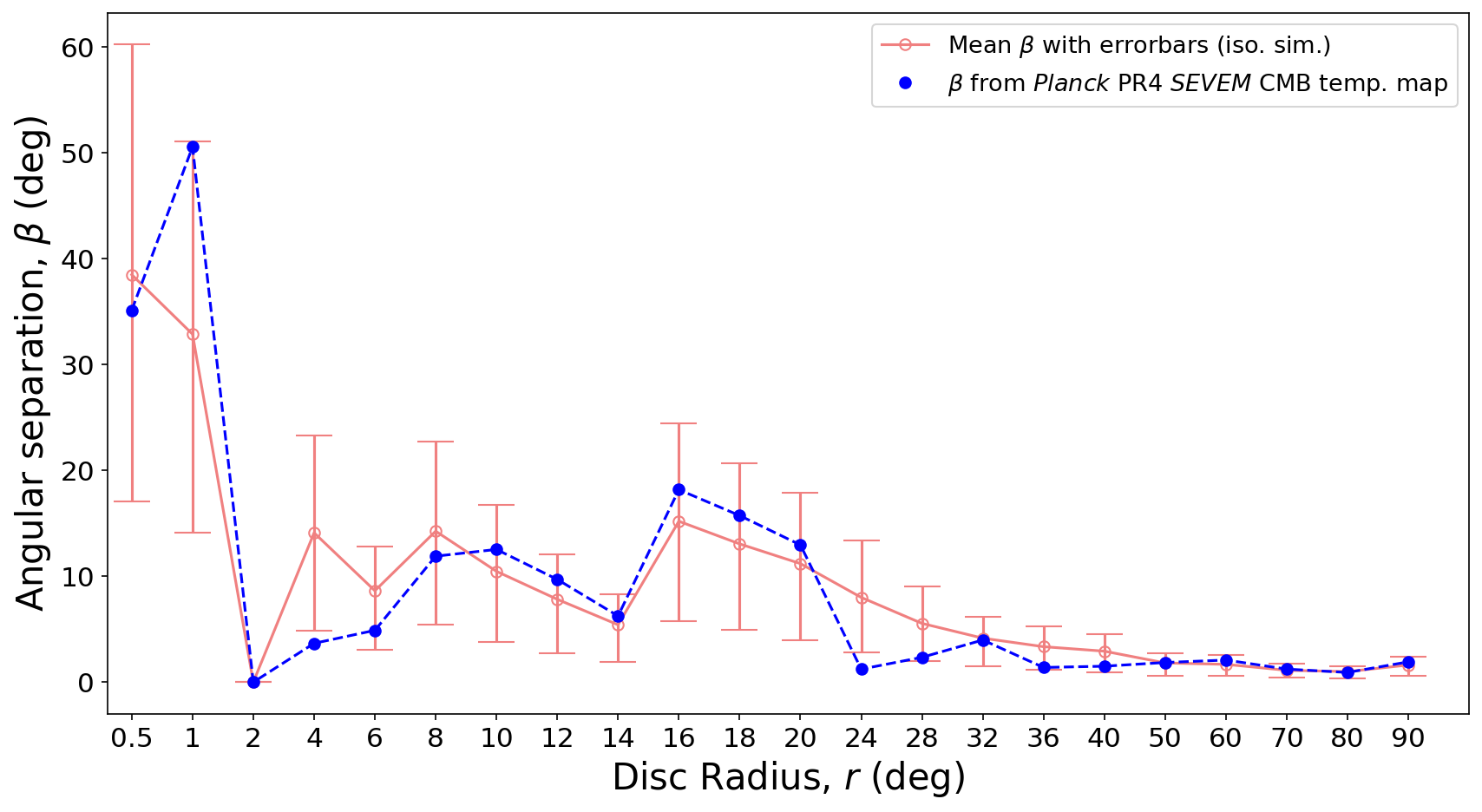}  
   \caption{Difference in the dipole directions recovered from LVE maps when employing fixed \nside=16 and varying \nside\ \healpix\ grids from data (\emph{blue} dashed line) and isotropic simulations (\emph{pink} solid line with $1\sigma$ scatter) are shown as a function of disc size `$r$'.}
   \label{fig:alpha-r-data}
\end{figure}

Dipole directions, determined from data LVE maps, are found to differ with the choice of disc radius and \nside\ of the LVM pixel grid. The directions so obtained are shifting away from galactic plane with increase in disc size in both cases. This is evident from the increasing circular point type size used to denote the recovered dipole directions in both cases. However this trend changes for $r \lesssim 2^\circ$ in data LVE maps derived at fixed \nside=16. In the other case (of varying \nside) the trend is consistent as stated.
To quantify this difference, let $\hat{d}$ and $\hat{e}$ be the dipole directions recovered from data LVE maps using fixed \nside=16 and varying \nside\ grids. Then the difference in the recovered dipole directions between the two schemes is given by $\beta = \cos^{-1}(\hat{d}\cdot\hat{e})$.
The distribution of these difference angles `$\beta$' (in degrees) in the recovered dipole directions from data and isotropic simulations are shown in Fig.~[\ref{fig:alpha-r-data}]. At $r\lesssim2^\circ$, the angular separation between the dipole directions from the two schemes is largest, while they are nearly zero for disc radius $r=24^\circ$ and beyond.

\begin{figure}
   \centering
   \includegraphics[width=0.98\columnwidth]{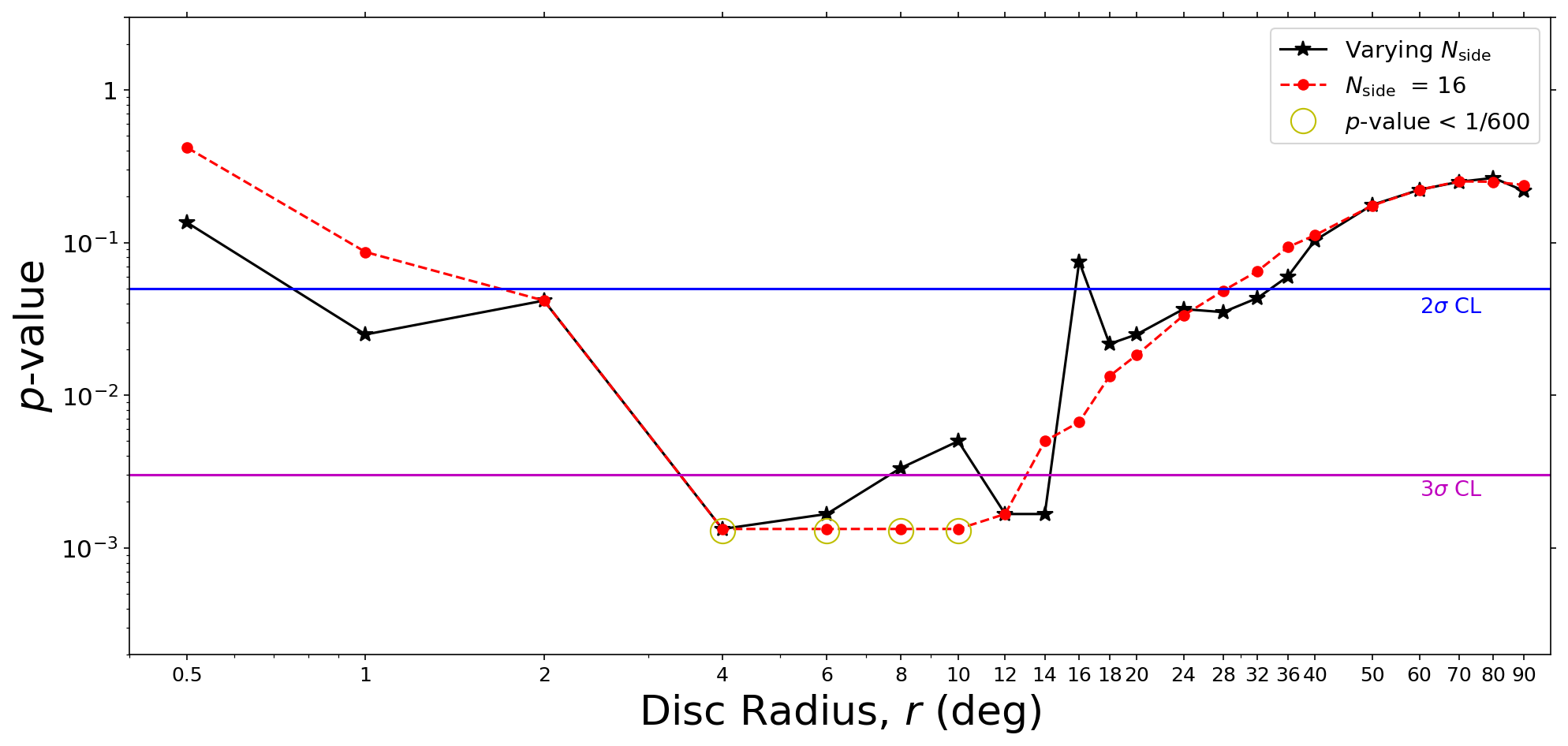}  
   \caption{$p$-values of the observed dipole amplitude in LVMs from \sevem\ 2020 CMB sky as a function of disc size `$r$' used to derive the LVE maps. \emph{Red} dots connected by a dashed line and \emph{black} stars joined by a solid line correspond to fixed \nside=16 and varying \nside\ schemes, respectively. An \emph{yellow} open circle denotes $p$-value $<1/600$ where there are no simulations whose LVMs' dipole amplitudes exceed the data value. The two horizontal lines indicate $2\sigma$ (\emph{blue}) and $3\sigma$ (\emph{magenta}) significances respectively.}
   \label{fig:p-value-dip-amp-data}
\end{figure}

The $p$-value plot of the dipole amplitudes recovered from data LVE maps for both cases of fixed \nside=16 and varying \nside\ are shown in Fig.~[\ref{fig:p-value-dip-amp-data}].
Here the $p$-values are being calculated in a frequentist manner i.e., as ratio of the number of times the dipole amplitude of LVE maps estimated from isotropic simulations exceeds that found in data maps to the total number of (isotropic) simulations used. In Fig.~[\ref{fig:p-value-dip-amp-data}] \emph{red} filled circle point types connected by a dashed line correspond to using fixed \nside=16 grid and \emph{black} stars connected by a solid line correspond to varying \nside\ grid in obtaining LVE maps. Some of the $p$-values are shown encircled by a larger \emph{yellow} open circle.
They highlight the disc sizes for which the $p$-values of dipole amplitude from LVE maps are zero per the (isotropic) simulation ensemble size used.
For large disc radii ($40^\circ$ onward) where the choice of \nside\ doesn't make any difference, $p$-values are same in both cases. In contrast, at lower disc radii the discrepancy is quite visible. For disc radius $4^\circ$ to $10^\circ$ in case of using fixed \nside=16, number of isotropic simulations (in terms of dipole amplitudes from corresponding LVE maps) exceeding that of data maps is zero implying that their random chance occurrence probability is $p<1/600\approx0.0016$ which is more than a $3\sigma$ significance. The analysis with varying \nside\ LVE maps indicate a significance of $\sim 3\sigma$ or less in the range of disc radii $4^\circ$ to $14^\circ$. It is clear from the plot that the LVE maps' dipole amplitudes seen in data are outside 95\% confidence level with either choice of the LVE map resolution for a wide range of disc sizes `$r$'.

The case with $r = 16^\circ$ and \nside=2 is somewhat different, as we do not have the freedom for smoothly transition between the map resolutions compared to disc sizes when using \healpix. From Table~\ref{tab:vary-nside-radii}, the length of diagonal of a pixel in an \nside=2 \healpix\  grid is $\approx41^\circ$. If circular discs are circumscribing the square shaped pixel then the radius of circular disc be at least $r=20^\circ$ to cover the full pixel information. But, for $r=16^\circ$ we are forced to omit some of the pixel information from the input map in computing LVM with \nside=2 \healpix\ grid. Instead, if we derive the LVE map at \nside=4 (whose diagonal-wise pixel size is $\approx21^\circ$ that leads to overlapping pixel information), we found that the $p$-value has more than $2\sigma$ significance.

\begingroup

\renewcommand{\arraystretch}{1.25} 
\begin{sidewaystable}
\centering
\begin{tabular}{| c | c | c | c | c | c | c | c | c |}
\hline
Disc Radius ($r$) & Fixed \nside\ & $A_{\rm LV,corr}$ & $(l_{\rm d},b_{\rm d})$ & $p$-value & Varying \nside\ & $A_{\rm LV,corr}$ & $(l_{\rm d},b_{\rm d})$ & $p$-value \\
\hline
0.5$^\circ$  & 16 &$0.008 \pm 0.013$&(213$^\circ$, -35$^\circ$) &0.4183 & 32  &$0.0093 \pm 0.0040$&(176$^\circ$,-22$^\circ$)&  0.1350\\
1$^\circ$  & 16 &$0.0236 \pm 0.0083$&(217$^\circ$, -67$^\circ$) &0.0866 & 32  &$0.0185 \pm 0.0050$&(187$^\circ$,-20$^\circ$)& \cellcolr 0.0250 \\
2$^\circ$  & 16 &$0.0187_{-0.0057}^{+0.0051}$& (194$^\circ$,-22$^\circ$)&\cellcolr 0.0416 & 16   &$0.0190 _{0.0057}^{0.0051}$ &(194$^\circ$,-22$^\circ$)& \cellcolr 0.0416 \\
4$^\circ$  & 16 &$ 0.0331 \pm 0.0060$& (205$^\circ$,-20$^\circ$)&\cellcolr $<$1/600 & 8 &$0.0341_{0.0064}^{+0.0065}$ &(207$^\circ$,-23$^\circ$)& \cellcolr $<$1/600 \\
6$^\circ$  & 16 &$0.0404 \pm 0.0070$& (209$^\circ$,-19$^\circ$)&\cellcolr $<$1/600 & 8 &$0.0360 \pm 0.0069$ &(204$^\circ$,-18$^\circ$)& \cellcolr 0.0016 \\
8$^\circ$  & 16 &$0.0450_{-0.0080}^{+0.0082}$& (210$^\circ$,-22$^\circ$)&\cellcolr $<$1/600 & 4 &$0.0420 _{-0.0090}^{+0.0089}$ &(198$^\circ$,-17$^\circ$)& \cellcolr 0.0033 \\
10$^\circ$ & 16 &$0.0467_{-0.0090}^{+0.0091}$& (213$^\circ$,-26$^\circ$)&\cellcolr $<$1/600 & 4 &$0.0407 \pm 0.0094$&(203$^\circ$,-35$^\circ$)& \cellcolr 0.0050 \\
12$^\circ$ & 16 &$0.0500_{-0.0098}^{+0.0096}$& (215$^\circ$,-29$^\circ$)&\cellcolr 0.0016 & 4   &$0.054_{-0.010}^{+0.011}$ &(223$^\circ$,-36$^\circ$)& \cellcolr 0.0016 \\
14$^\circ$ & 16 &$0.051\pm0.011$& (218$^\circ$,-31$^\circ$)&\cellcolr 0.0050 & 4   &$0.061\pm0.011$ &(225$^\circ$,-33$^\circ$)& \cellcolr 0.0016 \\
16$^\circ$ & 16 &$0.053\pm0.012$& (221$^\circ$,-34$^\circ$)&\cellcolr 0.0066 & 2   &$0.043_{-0.015}^{+0.016}$ &(233$^\circ$,-49$^\circ$)& 0.0750 \\
18$^\circ$ & 16 &$0.054\pm0.013$& (222$^\circ$,-36$^\circ$)&\cellcolr 0.0133 & 2   &$0.056\pm0.015$ &(231$^\circ$,-50$^\circ$)& \cellcolr 0.0216 \\
20$^\circ$ & 16 &$0.056_{-0.015}^{+0.013}$& (223$^\circ$,-37$^\circ$)&\cellcolr 0.0183 & 2   &$0.056 \pm 0.016$ &(232$^\circ$,-48$^\circ$)& \cellcolr 0.0250 \\
24$^\circ$ & 16 &$0.058_{-0.017}^{+0.015}$& (225$^\circ$,-38$^\circ$)&\cellcolr 0.0333 & 2   &$0.058\pm0.017$ &(227$^\circ$,-38$^\circ$)& \cellcolr 0.0366 \\
28$^\circ$ & 16 &$0.059_{-0.019}^{+0.017}$& (227$^\circ$,-39$^\circ$)&\cellcolr 0.0483 & 2   &$0.066_{-0.019}^{+0.018}$ &(224$^\circ$,-39$^\circ$)& \cellcolr 0.0350 \\
32$^\circ$ & 16 &$0.059_{-0.020}^{+0.018}$& (228$^\circ$,-40$^\circ$)&0.0650 & 2   &$0.068_{-0.020}^{+0.019}$ &(223$^\circ$,-38$^\circ$) & \cellcolr 0.0433 \\
36$^\circ$ & 16 &$0.056_{-0.021}^{+0.020}$& (230$^\circ$,-40$^\circ$)&0.0933 & 2   &$0.067_{-0.022}^{+0.019}$&(228$^\circ$,-40$^\circ$)&0.0600 \\
40$^\circ$ & 16 &$0.053_{-0.022}^{+0.020}$&(231$^\circ$,-41$^\circ$)&0.1116 & 2   &$0.056_{-0.022}^{+0.019}$&(230$^\circ$,-41$^\circ$)&0.1033 \\
50$^\circ$ & 16 &$0.042_{-0.024}^{+0.022}$&(235$^\circ$,-41$^\circ$)&0.1750 & 2   &$0.043_{-0.023}^{+0.022}$&(237$^\circ$,-42$^\circ$)&0.1766 \\
60$^\circ$ & 16 &$0.033_{-0.023}^{+0.022}$&(237$^\circ$,-42$^\circ$)&0.2216 & 2   &$0.035_{-0.023}^{+0.022}$&(234$^\circ$,-42$^\circ$)&0.2216 \\
70$^\circ$ & 16 &$0.030_{-0.021}^{+0.020}$&(238$^\circ$,-43$^\circ$)&0.2516 & 2   &$0.028_{-0.021}^{+0.019}$&(239$^\circ$,-43$^\circ$)&0.2500 \\
80$^\circ$ & 16 &$0.025_{-0.019}^{+0.017}$&(238$^\circ$,-44$^\circ$)&0.2500 & 2   &$0.022_{-0.019}^{+0.017}$&(239$^\circ$,-44$^\circ$)&0.2650 \\
90$^\circ$ & 16 &$0.022_{-0.016}^{+0.015}$&(239$^\circ$,-45$^\circ$)&0.2383 & 2   &$0.024_{-0.016}^{+0.015}$&(236$^\circ$,-45$^\circ$)&0.2183 \\
\hline
\end{tabular}
\caption{Bias corrected dipole amplitudes underlying LVE maps, $A_{\rm LV,corr}$, estimated at different disc radii from \sevem\ PR4 CMB map and the corresponding $p$-values are presented.
{ $p$-values highlighted in coloured cells denote disc sizes for which the observed HPA dipole amplitude is more than $2 \sigma$ significant}. Also listed are the corresponding dipole directions $(l_{\rm d},b_{\rm d})$ in galactic coordinates.}
\label{tab:A_corr-errbar}
\end{sidewaystable}

\endgroup

\begin{figure}
    \centering
    \includegraphics[width=0.95\linewidth]{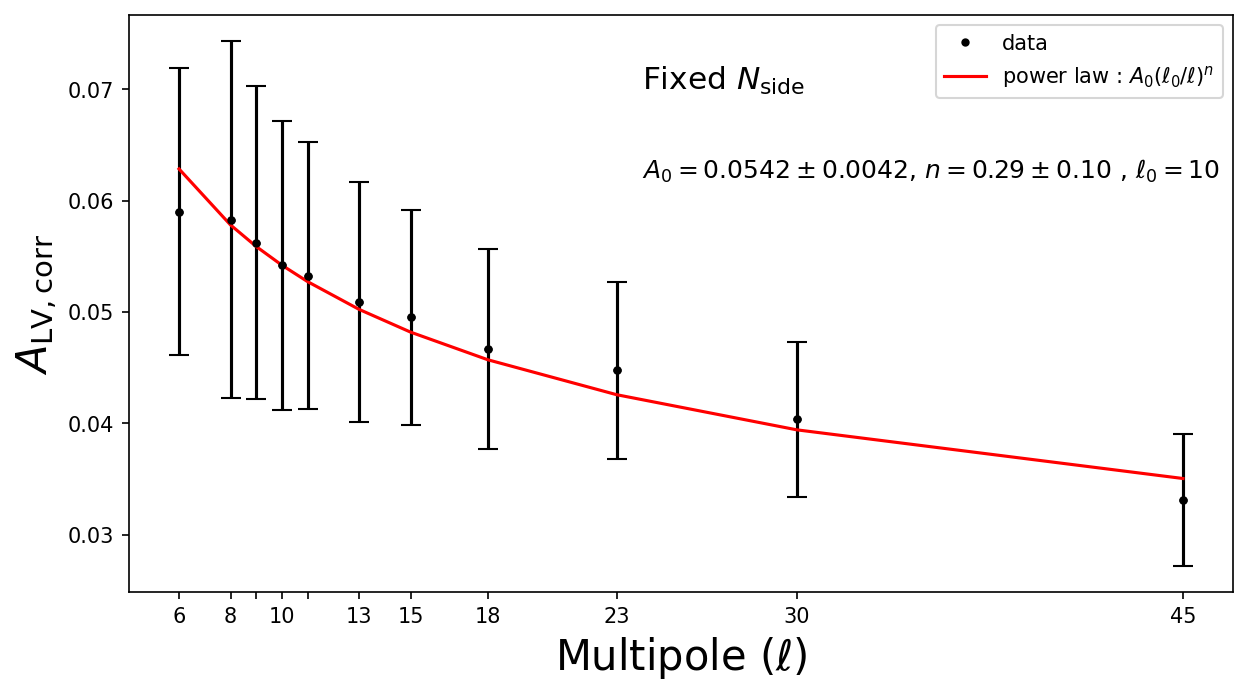}
    ~
    \includegraphics[width=0.95\textwidth]{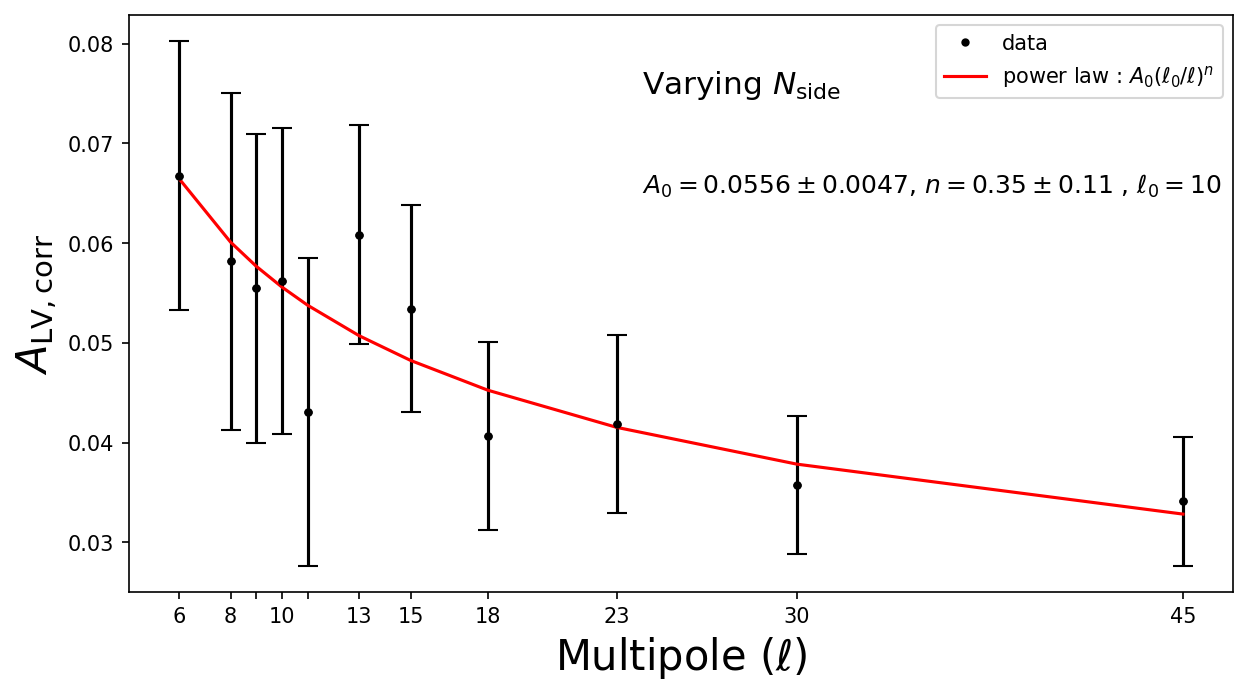}
    \caption{Power-law fits to the observed (bias corrected) LVE maps' dipole amplitudes. The power-law model clearly passes through most of the data points in both schemes of estimating LVM dipole amplitudes. Here we chose the pivot multipole to be $\ell_0=10$ for the power-law.}
    \label{fig:scale-dep-ns16-dns}
\end{figure}

In Table~\ref{tab:A_corr-errbar}, we summarize our findings viz., the dipole amplitudes, directions, and the corresponding $p$-values found for various disc radii chosen to compute LVE maps in the two cases of fixed \nside=16 and varying \nside\ grid from data.
LVE maps' dipole amplitudes that we recover are biased due to a random dipole component as per Eq.~(\ref{eq:dipole-expt-val}) and (\ref{eq:dipole-A}). Following Eq.~(\ref{eq:A_corr}) we obtain the bias corrected dipole amplitudes `$A_{\rm LV,corr}$', where the random dipole power is calculated from the isotropic simulations. These bias corrected LVM dipole amplitudes obtained from data are presented in Table~\ref{tab:A_corr-errbar} along with $1 \sigma$ error for different disc radii. Coloured cells in Table~\ref{tab:A_corr-errbar} correspond to disc radii whose $p$-values of the observed dipole amplitude in data LVE maps are significant by more than 2$\sigma$ CL.

From Table~\ref{tab:A_corr-errbar} (where bias corrected dipole amplitudes, $A_{\rm LV, corr}$ are presented) and also in Fig.~[\ref{fig:data-dip-ampl-error-ds}] (top panel), we see a clear dependence of these amplitude values with disc radius. So there is an indication for scale dependence of the LVM dipole amplitudes. To test this, we convert a disc radius `$r$' into `multipole' as $\ell=180^\circ/r$. With this conversion we test the scale dependence of dipole modulation amplitude by modeling it as an (inverse) power-law i.e.,
\begin{equation}
    A_{\rm LV,corr} = A(\ell) = A_{0}\left(\frac{\ell_0}{\ell}\right)^n
    \label{eq:hpa-powerlaw}
\end{equation}
where `$A_0$' is the amplitude of dipole modulation at some pivot multipole `$\ell_0$' and `$n$' is the spectral tilt (power-law index) in this model. If the dipole modulation is scale independent we will find $n=0$ (within error bars). We note that a power-law dependence of HPA dipole modulation amplitude was probed earlier, but in very few works~\cite{Aiola_2015,Tarun2019anom,Gonzalez2019hpal23}. Generally, this phenomenon is thought of as a step function with constant dipole amplitude ($\sim0.07$) up to multipoles $l\sim60$, after which the CMB sky is consistent with isotropy~\cite{Hoftuft_2009_lowL}. However, an angular clustering of HPA directions seem to be present up to a much higher multipole of $l\gtrsim600$~\cite{Hansen2009angclstr,plk2013isostat,plk2015isostat,plk2018isostat}.

{ We did an MCMC likelihood analysis for the power-law model in Eq.~(\ref{eq:hpa-powerlaw}). Details of MCMC fit are presented in Appendix~\ref{apdx:pl-fit-mcmc}.}
The power-law fits for data derived `$A_{\rm LV,corr}$' vs `$\ell$' are presented in Fig.~[\ref{fig:scale-dep-ns16-dns}] where the results from using a fixed \nside=16 LVM grid are depicted in the \emph{top} panel and the same from employing varying \nside\ LVM grid are presented in the \emph{bottom} panel, respectively. We clearly see that the power-law fits better passing through most of the data points in both cases. The amplitude and power-law index parameters are found to be $A_0=0.0542\pm0.0042$ and $n=0.29\pm0.10$ for the fixed \nside=16 case and $A_0=0.0556 \pm 0.0047$ and $n=0.35\pm 0.11$ when using varying \nside\ grid to compute LVMs for the pivot scale of $\ell_0=10$. Though the estimated amplitudes `$A_0$' at the chosen pivot multipole are essentially same in both cases, the power-law index is steeper in case of using varying \nside\ grid for LVE maps.
As is evident, in both cases, the amplitude of modulation decreases with increasing multipole `$\ell$' i.e., HPA is confined to only large angular scales of the CMB sky and {decreases} at small angular scales.

As mentioned above, few studies indeed considered scale dependence of hemispherical power asymmetry before. Using the values reported by Shaikh et al. (2019)~\cite{Tarun2019anom} (as quoted in their abstract : $A_{0} = 0.064 \pm 0.022$, $n = 0.92 \pm 0.22$) and  Marcos-Caballero et al. (2019)~\cite{Gonzalez2019hpal23} (as quoted in Table~2 of their paper : $A_{0} = 0.172 \pm 0.082$, $n = 0.59 \pm 0.35$) gives $A_{0} = 0.099 \pm 0.044$ and $0.114\pm0.027$ respectively at $\ell_0=10$. (We note that they used different pivot scales viz., $\ell_0=16$ and $5$ respectively. So we derived the rescaled values at $\ell_0=10$ to compare with our result.) These values are higher than the one we obtained (and, interestingly, also doesn't agree with each other after re-pivoting though they can be deemed consistent with each other within error bars). Recalling the relation between the variance and power spectrum of a CMB map :
\(
\langle \Delta T (\hat{n}) \rangle^2 = \sum_l (2l+1)C_l/4\pi
\),
we opine that part of the reason for this difference in the amplitude (and spectral index) estimates could be due to the presence of isotropic modes in variance computation up to a particular disc size `$r$' that dilute the HPA signal.

\section{Conclusions}
\label{sec:concl}
In this paper, we made a reassessment of Hemispherical Power Asymmetry (HPA) seen in CMB temperature data, first in NASA's WMAP 1yr CMB maps and then in all its subsequent data releases as well as all the data releases from ESA's \planck\ probe.
We made use of the \sevem\ CMB solution from the final data release (PR4) of \planck\ mission.
A set of 600 complementary simulations were provided as part of PR4 corresponding to observed CMB sky derived by employing \sevem\ cleaning method. Using the local variance estimator (LVE) method, we probed HPA as a dipole modulation of otherwise isotropic CMB sky. We employed a range of disc radii starting from $1^\circ$ to $90^\circ$ to map locally computed variances from the observed CMB sky. To understand the observed patterns better we extended the analysis further to $r=0.5^\circ$ disc radius. These local variances are computed from regions outside the PR3 common mask, recommended for analysis of CMB data from \planck's public release 4.

First, we re-validated the method using three sets of simulation ensembles viz., the isotropic CMB realizations with noise from \planck\ PR4 i.e., FFP12 simulation set as given (iso), then dipole modulated CMB maps (pure-dm), and low-$l$ only modulated CMB maps (low-l-dm) that are also added with corresponding FFP12 \sevem\ processed noise maps. Our purpose of re-evaluation is to explicitly demonstrate some aspects of the LVE map estimation procedure, specifically, the use of inverse variance weighting to fit the dipole underlying LVE maps, and the correlations that exist between different locally computed variances in LVE maps due to use of overlapping information (pixels) from input CMB map which are especially relevant when mapping local variances using larger disc radii. We have shown that when using a fixed \nside=16 for deriving LVE maps (whose \healpix\ pixel grid size is $\sim 4^\circ$), some parts of the CMB sky are not used when considering small disc radii such as $2^\circ$ or less, where as, some regions (pixels of CMB sky) are common to many locally computed variances in an LVE map when considering larger disc radii and consequently (very well) correlated.

In order to minimize the correlations in LVE maps to probe HPA, we showed that it can be achieved by matching the \nside\ of an LVE map with the disc radius chosen as closely as possible. So instead of using a fixed \nside=16 as originally used~\cite{Akrami2014}, we should use a varying \nside\ grid for generating LVE maps. Not only does it reduce the correlations in LVE maps as demonstrated in Fig.~[\ref{fig:cov-corr-matrix}], it also fastens the LVE maps' estimation process. Even after matching an LVE map's \nside\ grid size with the disc radius used, the variance of an LVE map's pixels are different. So using this information gives an improved fit to the estimation of dipole underlying an LVE map. We note that in our revalidation of LVE method, the difference is negligible, statistically, when fitting the dipole from LVE maps as they are and using the diagonal elements of the LVE maps' covariance matrix for a particular disc radius from simulations as weights (inverse variance weighting). This could be because of the strong signal to noise ratio of the CMB temperature data.
However there is a variation in the dipole information recovered at the individual map level and can consequently affect the significance of the underlying dipole modulation amplitude in observations.
We expect that these choices will be more relevant to CMB polarization data which will be undertaken in a future work. We also presented how to correctly estimate the bias corrected LVM dipole amplitudes (Eq.~\ref{eq:A_corr}) unlike the simple subtraction (correction) done at the amplitude level.

Some of the observations made during this re-evaluation of LVE method are as follows. In our study we simulated dipole modulated maps with an amplitude of $A_{\rm d}=0.072$ in the direction $(l_{\rm d},b_{\rm d})=(221^\circ,-20^\circ)$. The amplitude of the dipole recovered from LVE maps were lower than the injected value for larger disc radii. This is true of both approaches of fixed \nside=16 and varying \nside\ to map local variances of a CMB sky, indicating a limitation on the reliability of the LVE method up to only some maximum disc radius. With a tolerance level of a $10\%$ deviation of the mean recovered dipole amplitude from the simulations, but nevertheless consistent with the injected dipole amplitude in them {within the standard deviation of the simulations, represented by error bars in the plot}, the LVE method can be used reliably up to a maximum disc radius of $\sim40^\circ$ as demonstrated in Fig.~[\ref{fig:pure-dm_err_bar}].
We also tested the range of validity i.e., the maximum disc radius up to which we can use the LVE method reliably vis-a-vis the strength of the underlying anisotropy. We found that for higher levels of dipole modulation amplitude, if present in a CMB map, the maximum disc radius to use is relatively lower. In any case, for very large disc radius the LVE method breaks down and should not be used.

After this comprehensive re-validation, we applied LVE method with fixed \nside=16 and varying \nside\ grids for different choices of disc radii from $r=0.5^\circ$ to $90^\circ$ on data. We used the more appropriate inverse variance weighting method to fit the dipole in LVE maps thus derived from \sevem\ cleaned CMB map from \planck\ PR4. We find that there is a broad clustering of the dipole directions underlying the observed CMB sky for different disc sizes chosen. Nevertheless, there is a tendency for the LVE maps' dipole directions to move away from the galactic plane with increasing disc radius. We further found that there is a difference in the dipole directions recovered between the two schemes employed, particularly at the low disc radii, the largest being $\sim 50^\circ$ for $r=1^\circ$.

The dipole amplitude estimated from the two approaches i.e., fixed \nside=16 and varying \nside\ for LVE maps gave slightly different amplitudes, expectedly, and so their $p$-values are also different. With fixed \nside=16 LVE maps, no simulation (out of 600) was found to have a dipole amplitude greater than that seen in data LVMs for disc sizes $r=4^\circ$ to $10^\circ$. With varying \nside, such a situation was seen only for $r=4^\circ$ when mapping the local variances using \nside=8 \healpix grid. 
Nevertheless, in case of fixed \nside=16 LVE maps, the dipole amplitudes are anomalous at $2\sigma$ confidence level or more for $r=2^\circ$ to $28^\circ$ disc radii.
When employing varying \nside\ grid for LVE maps, for disc radius of $1^\circ$ to $32^\circ$ (except for $r=16^\circ$) the estimated dipole amplitudes were anomalous at $2\sigma$ significance or better.

Thus, we confirm earlier findings that there is a significant hemispherical power asymmetry underlying the CMB sky as a dipole modulation field. We also confirm that the dipole direction of HPA is consistent, broadly, when assessing with different disc sizes. Since different modes (angular scales in a CMB map) contribute to estimating variances locally when using different disc sizes, the dipole directions thus recovered could in principle be different. So we are driven to conclude that the HPA direction is robust with respect to different choices of disc radii (or equivalently filter sizes) as reported in earlier studies. Nevertheless the dipole amplitude is different for different disc radii used, implying that we have a scale dependent HPA amplitude.
Modeling the observed dipole amplitudes as a power-law with angular size viz., $A_{\rm LV,corr} = A_0(\ell_0/\ell)^n$, where $\ell=180^\circ/r$, we find that $A_0\sim 0.05$ in both cases of using fixed \nside=16 or varying \nside\ schemes for LVM estimation and $n>0$ implying that the dipole modulation amplitude is scale dependent.
Therefore, HPA is significant at large angular scales that decreases in strength at small scales of the CMB sky. 
So, the presence of hemispherical power asymmetry in CMB data remains as an open question within the framework of standard cosmological model.

\section*{Acknowledgments}
SS acknowledges Ph.D. fellowship received from the University Grants Commission (UGC), India through UGC-Ref No.:1487/(CSIR-UGC NET JUNE 2018). A.S. would like to acknowledge the support by National Research Foundation of Korea 2021M3F7A1082056, and the support of the Korea Institute for Advanced Study (KIAS) grant funded by the government of Korea.
Part of the results presented here are based on observations obtained with \planck\, an ESA science mission with instruments and contributions directly funded by ESA Member States, NASA, and Canada.
This research used resources of the National Energy Research Scientific Computing Center (NERSC), a U.S. Department of Energy Office of Science User Facility operated under Contract No. DE-AC02-05CH11231.
Further, this work also made use of the following software packages :
{\tt HEALPix}/{\tt Healpy}~\cite{healpix,healpy},
\texttt{SciPy}\footnote{\url{https://scipy.org}}~\cite{scipy2020},
\texttt{NumPy}\footnote{\url{https://numpy.org}}~\cite{numpy2020},
\texttt{Astropy}\footnote{\url{http://www.astropy.org}}~\cite{astropy2013,astropy2018,astropy2022},
and \texttt{matplotlib}\footnote{\url{https://matplotlib.org/stable/index.html}}~\cite{matplotlib}. Finally, we thank the anonymous referee/Editor for many useful comments that helped improve and bring more clarity to the presentation of this work. We further want to thank the referee for pointing out few more (minor) issues to investigate in the future.

\bibliographystyle{unsrt}
\bibliography{ref_hpa_plk20_etal}

\appendix
\counterwithin{figure}{section}
\counterwithin{table}{section}

\section{LVE maps' dipole estimation with simple and inverse variance weighting schemes}
\label{apdx:norm-inv-wght-dip-fit}

Here we dwell on few details regarding estimation of dipole underlying normalized local variance maps that were commented about in earlier works, though not explicitly demonstrated.

We use the \verb+remove_dipole+ functionality of \healpix\ as it is to estimate dipole from normalized LVE maps from data or simulations which we call as \emph{simple fit}. When using a mask, depending on the number of pixels available at a specific location at which disc centers are defined to compute local variances, the local variances computed will differ. Further the variances themselves follow a $\chi^2$ distribution as the CMB temperature anisotropies are expectantly Gaussian random fluctuations on the sky. Thus the variance of individual pixels in an LVE map differs. Therefore we would get a better fit for the dipole, if we employ a weighted fit to estimate the dipole from normalized LVE maps. For this \emph{inverse variance weighted fit}, we use the same \healpix's \verb+remove_dipole+ function but will now make use of the inverse of the diagonal elements of the corresponding covariance matrix per Eq.~(\ref{eq:cov-mat}) as weights to get dipole estimates.

In Sec.~\ref{sec:vald}, we have already shown the covariance matrices and the corresponding correlation matrices that highlights the need for a weighted fitting using the diagonal elements of the covariance matrix for a particular choice of disc radius (in Fig.~[\ref{fig:cov-corr-matrix}]). We have also shown that using different \nside\ for deriving LVE maps for different choices of disc sizes, there will be minimum to no correlations between pixels of an LVE map. We also do the same for the fixed \nside=16 case in our comparison that follows.

\begin{figure}
\centering
\includegraphics[width=0.9\textwidth]{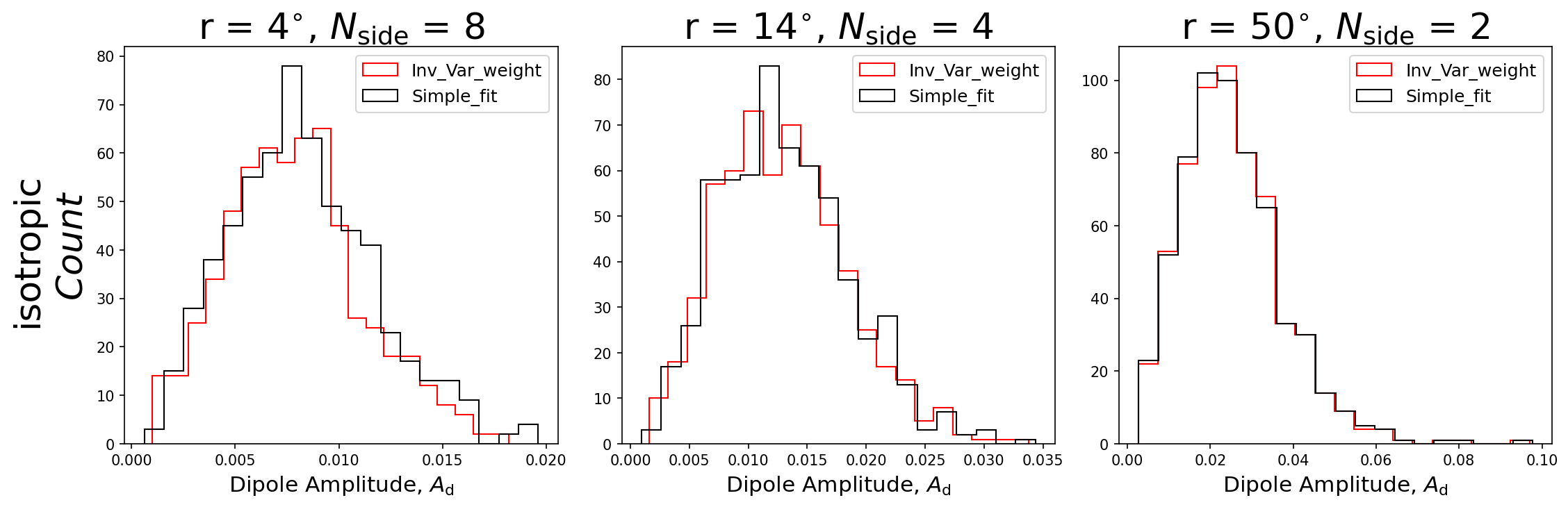}
~
\includegraphics[width=0.9\textwidth]{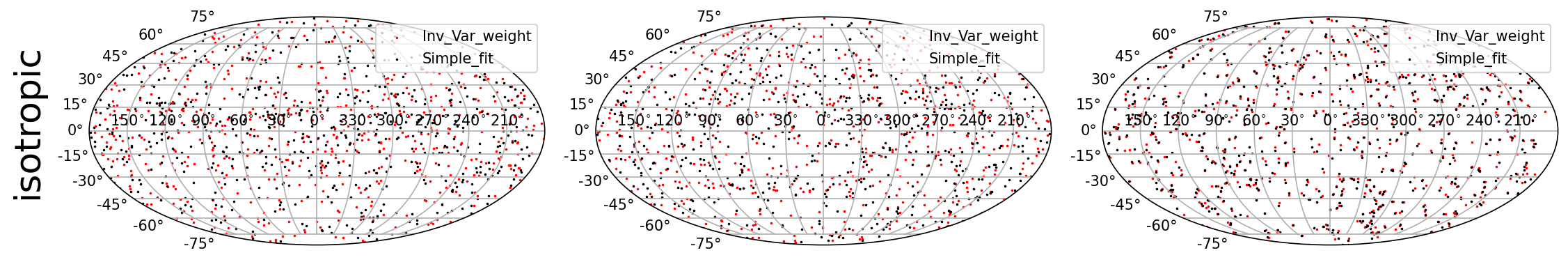}
~
\includegraphics[width=0.9\textwidth]{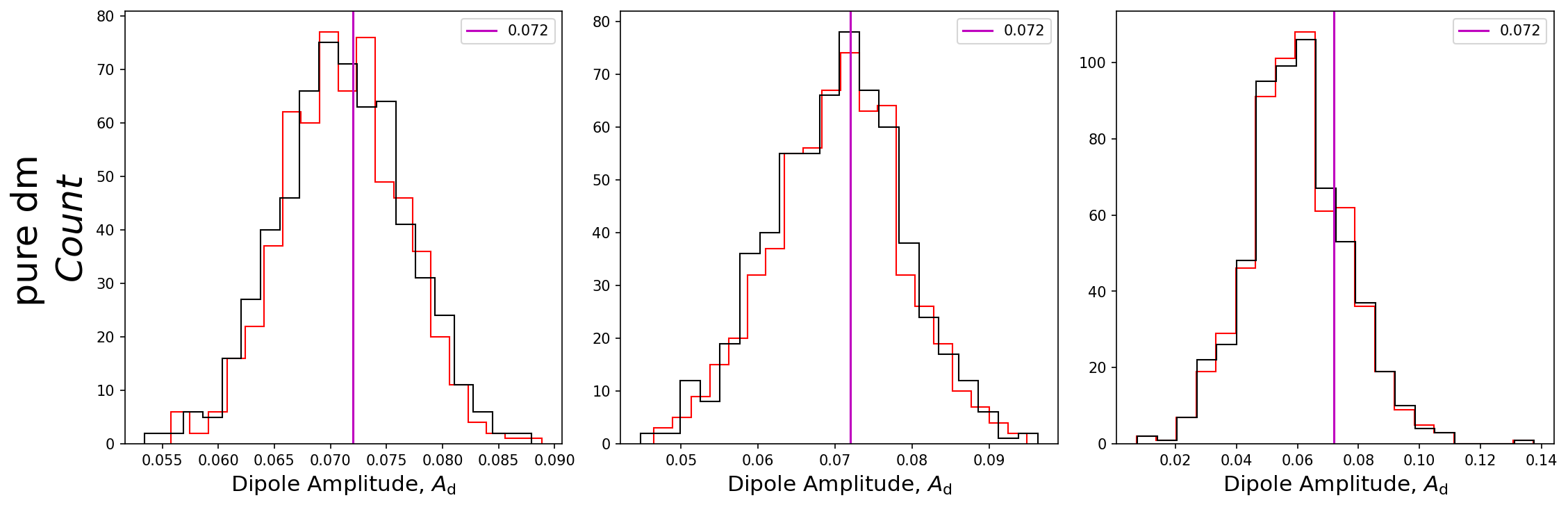}
~
\includegraphics[width=0.9\textwidth]{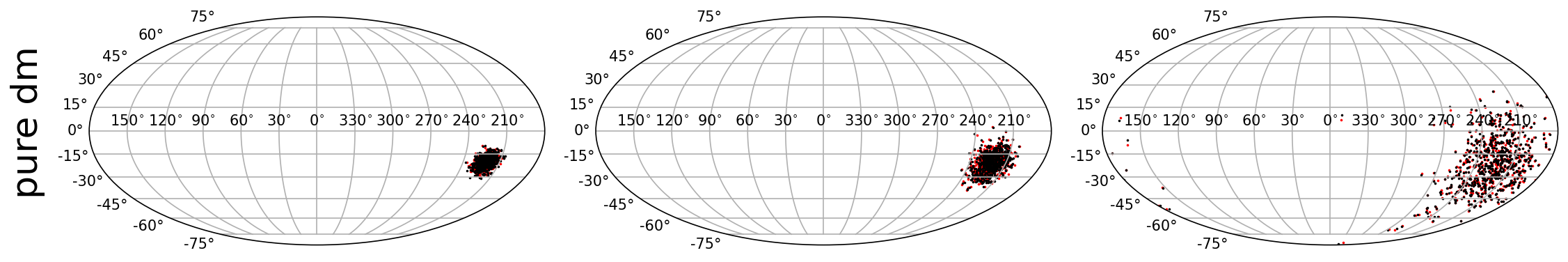}
~
\includegraphics[width=0.9\textwidth]{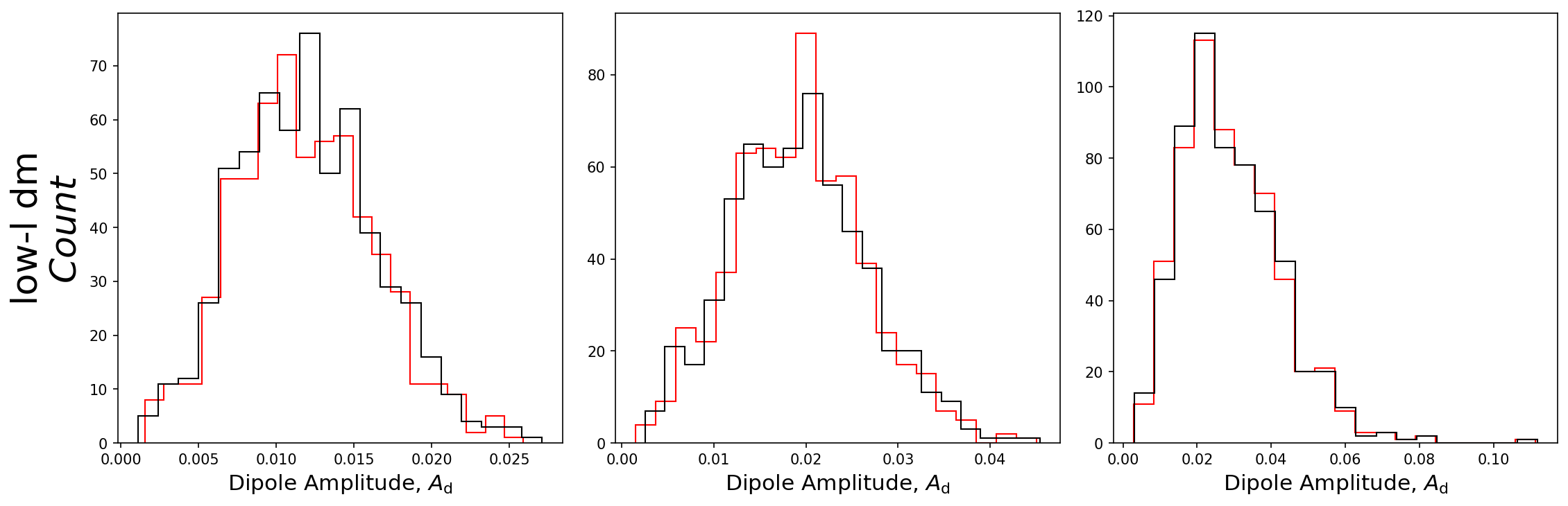}
~
\includegraphics[width=0.9\textwidth]{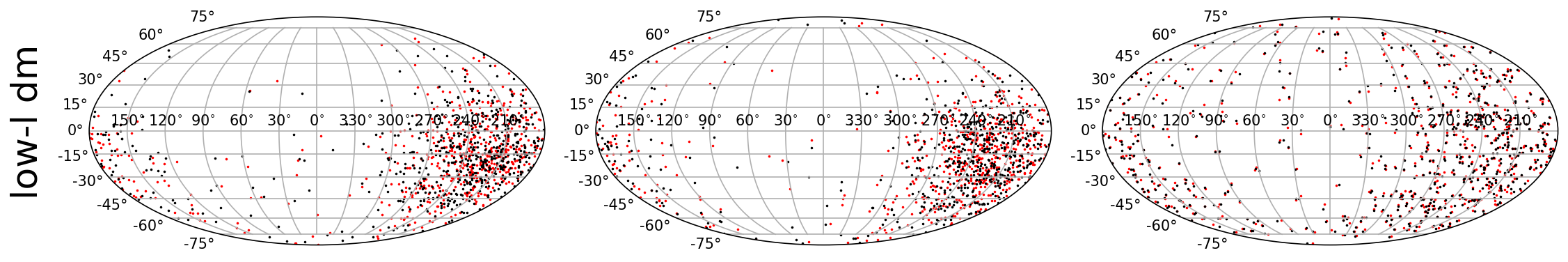}
\caption{For some select disc radii $r=4^\circ$, $14^\circ$ and
         $50^\circ$, distribution of recovered dipole amplitude ($A_{\rm d} = A_{\rm LV}/2$) and direction are shown when
         using a simple fit (in \emph{black}) and inverse variance weighting (in \emph{red})
         schemes, where varying \nside\ grid is used in deriving LVE maps.
         The \emph{first} and \emph{second} rows
         correspond to results from isotropic simulation set, \emph{third} and \emph{fourth}
         correspond to results from pure dipole modulated (pure dm) simulations, and
         finally the \emph{fifth} and \emph{sixth} correspond to recovered dipole amplitude and
         direction respectively from low-$l$ only modulated (low-$l$ dm) simulations.}
\label{fig:apdx:norm-inv-wght-dip-fit-varyns}
\end{figure}

\begin{figure}
\centering
\includegraphics[width=0.92\textwidth]{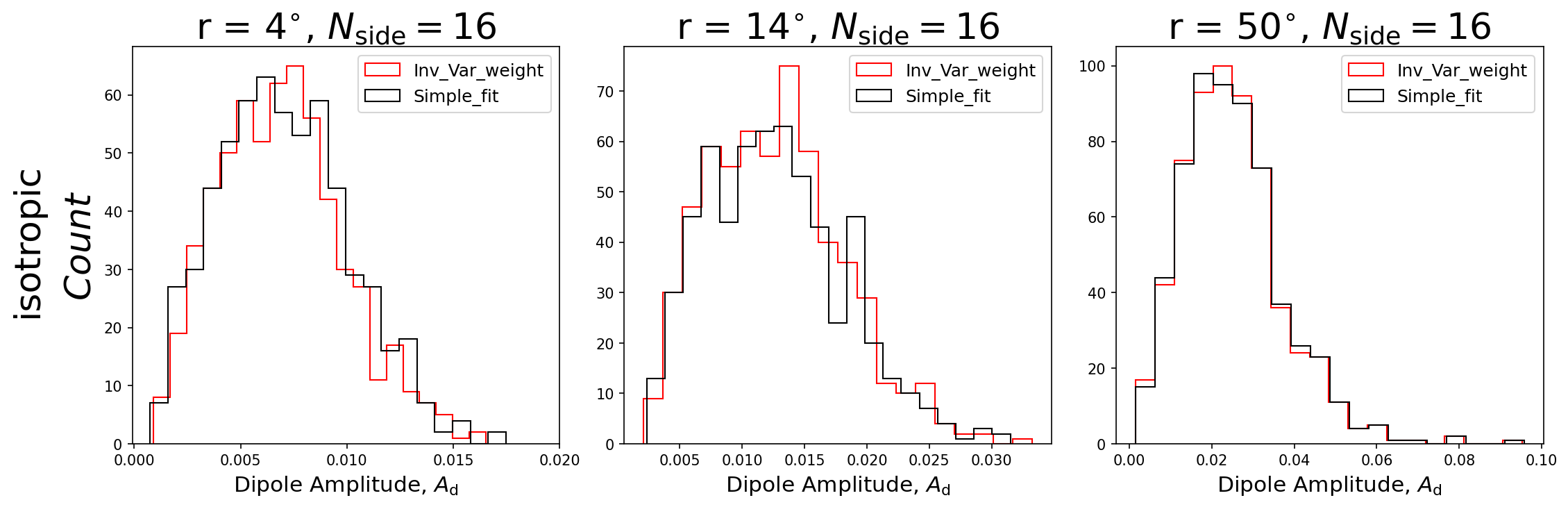}
~
\includegraphics[width=0.92\textwidth]{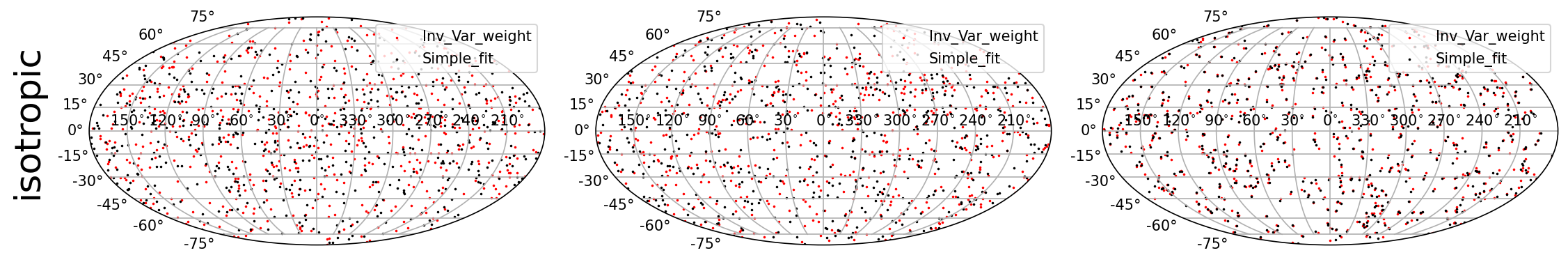}
~
\includegraphics[width=0.92\textwidth]{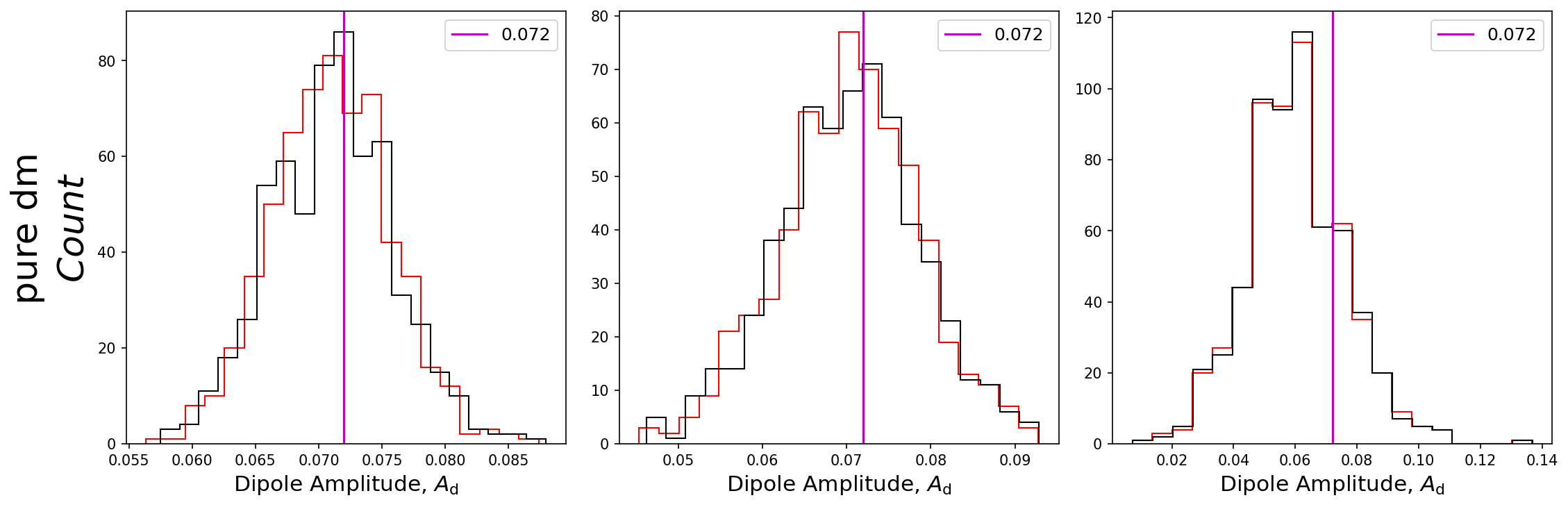}
~
\includegraphics[width=0.92\textwidth]{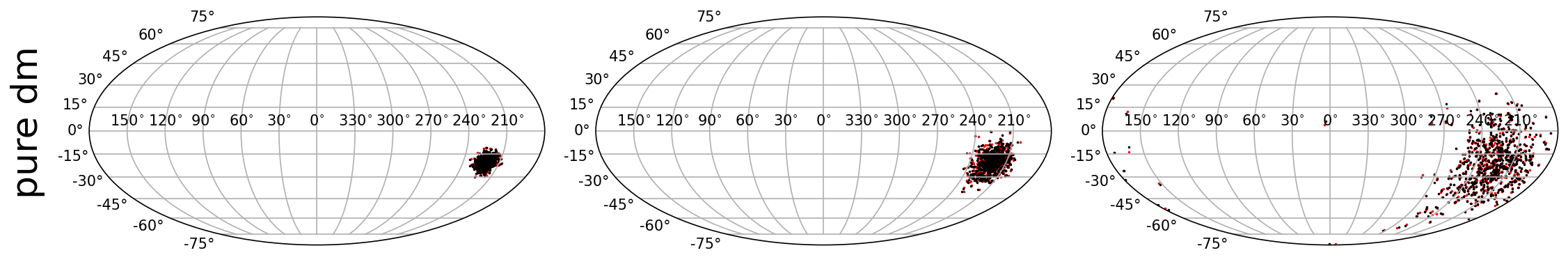}
~
\includegraphics[width=0.92\textwidth]{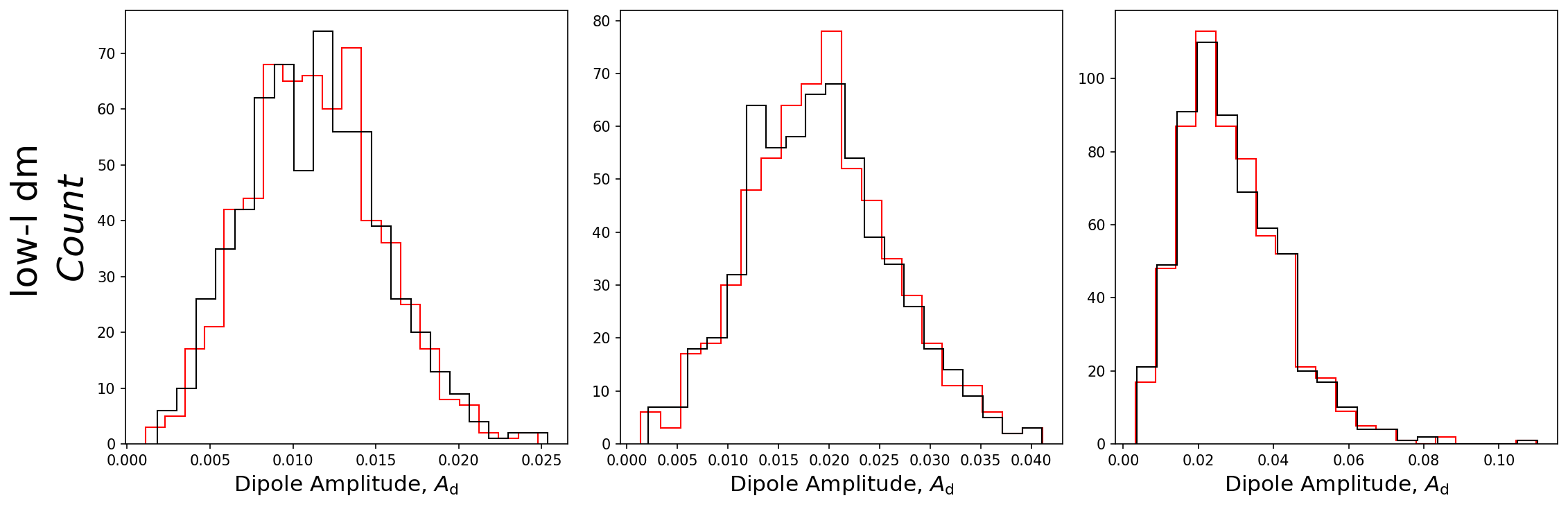}
~
\includegraphics[width=0.92\textwidth]{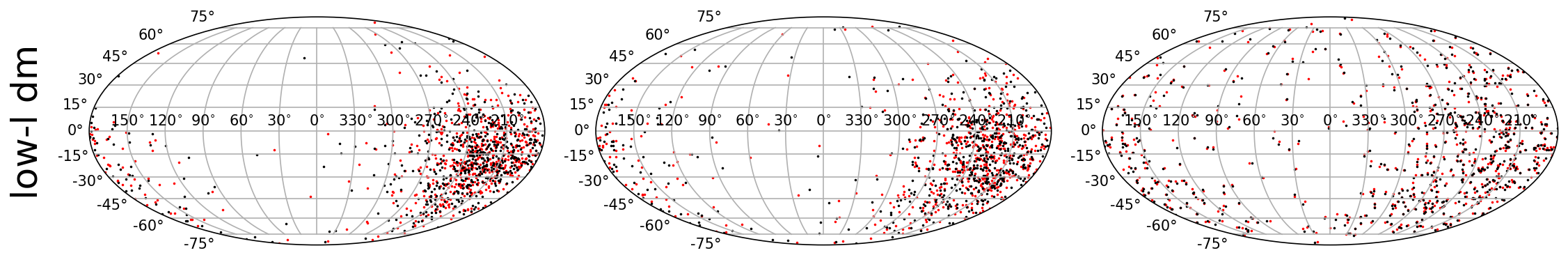}
\caption{Same as Fig.~[{\ref{fig:apdx:norm-inv-wght-dip-fit-varyns}}], but when using
         fixed \nside\ grid to derive LVE maps.}
\label{fig:apdx:norm-inv-wght-dip-fit-ns16}
\end{figure}

\begin{figure}[t]
\centering
\includegraphics[width=0.88\textwidth]{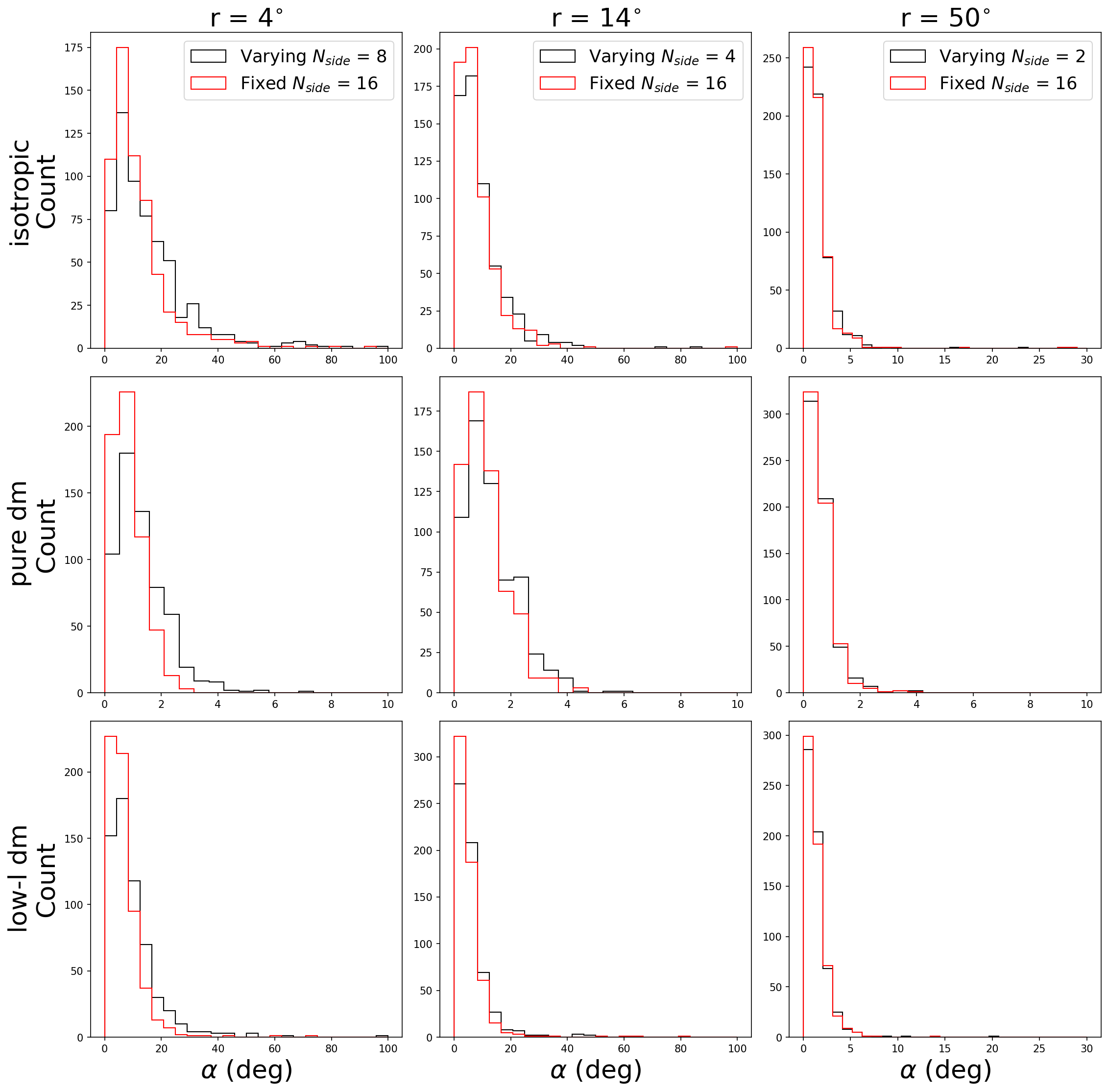}
\caption{Histogram of angular separation between dipole directions recovered from LVE maps as estimated using  a simple fit and an inverse variance weighting in varying \nside\ (\emph{black}) or fixed \nside=16 (\emph{red}) cases for some select disc sizes.}
\label{fig:apdx:sim-inv-var-alpha}
\end{figure}

In Fig.~[\ref{fig:apdx:norm-inv-wght-dip-fit-varyns}], we show the recovered dipole amplitudes and directions for disc radii $4^\circ$, $14^\circ$, and $50^\circ$ respectively in the first, second and third columns from isotropic simulations (first and second rows), pure dipole modulated simulations (third and fourth rows) and low multipole only dipole modulated maps (fifth and sixth rows).
Each plot in that figure depicts results that were obtained following the two dipole fitting schemes described above. \emph{Red} colour is used for fixed \nside=16 case and \emph{black} is used for varying \nside\ case. The vertical line in \emph{magenta} with $A_{\rm d}=0.072$ represents the injected dipole amplitude in simulations. As such there is no visible difference in the directions and amplitudes of the recovered dipole from normalized LVE maps using the two schemes. This could be because the diagonal elements of the covariance matrices of LVE maps for any disc radius are not very different. Indeed, from their histograms, we find that most of the diagonal elements are located around the same value or have nearly same order of magnitude. Hence a simple fit equivalent to a least square fit is resulting in similar estimates for LVE maps' dipoles in comparison to using inverse variance weighting. This in turn could be a consequence of the fact that CMB temperature maps are signal dominated. However this might be very relevant in case of polarization analysis where even the latest full sky measurements from \planck\ are noisy. In Fig.~[\ref{fig:apdx:norm-inv-wght-dip-fit-ns16}] we present the same results but for the fixed \nside=16 case. Our observations from the varying \nside\ case remain the same here also.

Now, in order to understand the difference in the dipole direction recovered when using simple fitting ($\hat{d}_s$) verses inverse variance weighting fit ($\hat{d}_w$), we compute the angular separation between the two  dipole directions thus recovered from a particular simulation as
$
\alpha = \cos^{-1}(\hat{d}_s \cdot \hat{d}_w)
$.
We do so in both cases of fixed \nside=16 and varying \nside\ LVE maps. The results are presented in Fig.~[\ref{fig:apdx:sim-inv-var-alpha}] as found in isotropic (top row),
pure dipole modulated (middle row) and low-$l$ dipole modulated (bottom row) CMB maps.
Column-wise, the histograms of angular separation between dipole directions from a simple fit and inverse variance weighting fit are shown for some select disc radii viz., $r=4^\circ$ (first column), $14^\circ$ (second column) and $50^\circ$ (third column). This angular separation `$\alpha$' between the two schemes to fit the dipole from LVE maps for both cases of varying \nside\ (in \emph{black} colour) and fixed \nside=16 (in \emph{red} colour) are exhibited in each subplot. It is obvious from the figure that in most cases we recover the same direction whichever dipole fitting scheme is employed either from isotropic or modulated simulations. Again this may be a consequence of the signal dominated CMB temperature data. This aspect should however be tested further in case of polarization.

With these in mind, in this work, we do the inverse variance fitting of LVE maps to estimate the underlying dipole using the diagonal elements of covariance matrix corresponding to that choice of disc radius `$r$' and \nside.

\section{MCMC fit to test scale dependence of HPA}
\label{apdx:pl-fit-mcmc}

To fit the power-law, we do an MCMC likelihood analysis. The error bars estimated from isotropic simulations were used in $\chi^2$ minimization as described below. ({We didn't consider covariance between the dipole amplitudes from different `$r$' as the diagonal elements of the covariance matrix are larger. So the rest are ignored as a reasonable approximation.}) We took disc sizes $r=4^\circ$ to $32^\circ$ in the power-law fitting of the LVE maps' dipole amplitudes as a function of angular scale. We first determined the multipole values ($\ell$) that correspond to each disc size `$r$' using the relation $\ell \sim 180^\circ/r$ and \emph{rounding} the resulting values. This rounding process produces degenerate `$\ell$' values. In such cases, the amplitude and error for each degenerate multipole `$\ell$' are calculated as the inverse variance weighted average of the amplitudes with symmetrized errors (obtained by taking average of upper and lower errors) from the corresponding disc sizes.

Mathematically, let the bias corrected amplitude of dipole from an LVE map for a particular disc radius is $A^{+\sigma_1}_{-\sigma_2}$. Then, we first get symmetrized error for that disc radius as $\sigma=(\sigma_1+\sigma_2)/2$. If two disc sizes `$r1$' and `$r2$' get rounded to the same `$\ell$', then we combine them using inverse variance weighting as follows:
\[
A_{\ell} = \frac{ A_{r1}/ \sigma_{r1}^2 + A_{r2} / \sigma_{r2}^2 }{ 1/ \sigma_{r1}^2 + 1/\sigma_{r2}^2 }  \quad {\rm and} \quad \sigma_\ell^2 = \frac{ 1 }{ { 1/ \sigma_{r1}^2 + 1/\sigma_{r2}^2 } }\,.
\]

In this way, we have 11 distinct multipoles $\ell = \{6, 8, 9, 10, 11, 13, 15, 18, 23, 30, 45\}$, the corresponding bias corrected dipole amplitudes `$A_\ell$', and associated error estimates `$\sigma_\ell$'. Using MCMC, we now fit for a power-law by minimizing the $\chi^2$ function that is defined as,
\[
\chi^2(A_0,n) = -2\ln[\mathcal{L}(A_0,n)] = \sum_\ell \frac{\left[A_\ell^{\rm data} - A_0({\ell_0}/{\ell})^n\right]^2}{\sigma_\ell^2}\,,
\]
where `$\mathcal{L}$' is the likelihood function, and $\ell_0=10$ is the pivot scale chosen.  Priors used for the fit parameters are: $A_0 = [0,0.2]$ and $n=[-2,2]$.

\end{document}